\documentclass[final,5p,twocolumn]{elsarticle}

\usepackage{lineno}

\usepackage{amsmath,amssymb,bm,subfigure,color,url,booktabs,mathrsfs}%cite
\usepackage[colorlinks=true,citecolor=blue,linkcolor=blue,urlcolor=blue]{hyperref}

\graphicspath{{fig/}}

\newcommand{\argmax}{\mathop{\rm argmax}\limits}
\definecolor{amazonite}{RGB}{0,115,150}
\definecolor{myred}{RGB}{255,56,0}
\definecolor{mygreen}{RGB}{30,150,30}
\definecolor{mybrown}{RGB}{150,30,30}
\definecolor{darkblue}{RGB}{30,50,100}
\definecolor{darkred}{RGB}{80,30,30}

\journal{Information Sciences}

\begin{document}

\begin{frontmatter}



\title{
Musical Rhythm Transcription Based on Bayesian Piece-Specific Score Models Capturing Repetitions\tnoteref{label1}
}
\tnotetext[label1]{This work is in part supported by JSPS KAKENHI Nos.\ 15K16054, 16H01744, 16J05486, and 19K20340; JST ACCEL No.\ JPMJAC1602; the Kyoto University Foundation; and the Kayamori Foundation. The work of EN was supported by the JSPS research fellowship (PD).}

\author[alabel1,alabel2]{Eita Nakamura\corref{cor1}}
\ead{eita.nakamura@i.kyoto-u.ac.jp}
\author[alabel1,alabel3]{Kazuyoshi Yoshii}
\cortext[cor1]{Corresponding author}
\address[alabel1]{Graduate School of Informatics, Kyoto University, Kyoto 606-8501, Japan}
\address[alabel2]{The Hakubi Center for Advanced Research, Kyoto University, Kyoto 606-8501, Japan}
\address[alabel3]{AIP Center for Advanced Intelligence Project, RIKEN, Tokyo 103-0027, Japan}

\begin{abstract}
Most work on musical score models (a.k.a.\ musical language models) for music transcription has focused on describing the local sequential dependence of notes in musical scores and failed to capture their global repetitive structure, which can be a useful guide for transcribing music. Focusing on rhythm, we formulate several classes of Bayesian Markov models of musical scores that describe repetitions indirectly using the sparse transition probabilities of notes or note patterns. This enables us to construct piece-specific models for unseen scores with an unfixed repetitive structure and to derive tractable inference algorithms. Moreover, to describe approximate repetitions, we explicitly incorporate a process for modifying the repeated notes/note patterns. We apply these models as prior musical score models for rhythm transcription, where piece-specific score models are inferred from performed MIDI data by Bayesian learning, in contrast to the conventional supervised construction of score models. Evaluations using the vocal melodies of popular music showed that the Bayesian models improved the transcription accuracy for most of the tested model types, indicating the universal efficacy of the proposed approach. Moreover, we found an effective data representation for modelling rhythms that maximizes the transcription accuracy and computational efficiency.
\end{abstract}


% \begin{highlights}
% \end{highlights}


\end{frontmatter}

% \linenumbers

%%%%%%%%%%%%%%%%%%%%%%%%%%%%%%%%%%%%%%%%%%%%%%
\section{Introduction}
\label{sec:Intro}
%%%%%%%%%%%%%%%%%%%%%%%%%%%%%%%%%%%%%%%%%%%%%%

Music transcription is an actively studied but yet unsolved problem in music information processing \cite{Benetos2019,Benetos2013}.
One of the goals of music transcription is to convert a musical performance signal into a human-readable symbolic musical score.
While recent studies have achieved highly accurate pitch detection methods \cite{Benetos2015,Hawthorne2018,Wu2019}, it is also necessary to transcribe rhythms in order to obtain symbolic music representations \cite{Cemgil2000,Desain1989,Hamanaka2003,LonguetHiggins1987,Nakamura2018ICASSP,Nakamura2016,Nakamura2017,Nishikimi2016,Raphael2002,Takeda2002,Temperley1999}.
Since there are many logically possible representations of rhythms (including inappropriate ones for transcription) for a given performance \cite{Cemgil2000}, using a (musical) score model (a.k.a.\ musical language model) that describes the prior knowledge of musical scores is a key to solving this problem.
Here, we study the problem of the rhythm transcription of monophonic music, which is the task of recognizing score-notated musical rhythms from human-performed MIDI data that contain onset time deviations.
A rhythm transcription method can also be used as a module of audio-to-score music transcription systems, as studied in \cite{Nakamura2018ICASSP,Nishikimi2016}.

A common approach for music transcription is to integrate a score model and a performance/acoustic model to obtain a proper transcription that best fits an input performance signal, similar to the statistical speech recognition method \cite{Rabiner1989}.
%More recently, end-to-end approaches have also been attempted \cite{Carvalho2017,Roman2018}, which have been of limited success so far.
This is a top-down approach for describing the generative process of musical performance signals using statistical models, in contrast to bottom-up approaches for extracting features from musical performance signals, such as using ratios of inter-onset intervals \cite{Jacoby2017,Ravignani2018} and using the wavelet transform \cite{Smith2008} to represent rhythmic features.
A major advantage of the former approach is the potential to utilize statistical machine learning techniques.

In constructing a score model for music transcription, capturing both local and global features of musical scores is considered to be effective.
Among the local features, the sequential dependences of musical notes can be used to induce output scores that obey the musical grammar.
Among the global features, repetitive structures are commonly found in various styles of music \cite{SweetAnticipation,Savage2015} and can be a useful guide for transcription since they can complement the local information of musical scores.
This is because by using a repetitive structure, one can in effect cancel out timing deviations and other ``noises'' in performances.
Conventional score models for music transcription that are designed to capture the local sequential dependence of musical notes include Markov models \cite{Hamanaka2003,Nakamura2017,Nishikimi2016,Raczynski2013,Raphael2002,Raphael2005,Takeda2002}, hidden Markov models (HMMs) \cite{Schramm2017}, recurrent neural networks (RNNs) \cite{Sigtia2016}, and long-short term memory (LSTM) networks \cite{Ycart2017}.
However, it is challenging to construct a score model incorporating a repetitive structure for transcription, particularly because the computational costs for imposing extensive constraints on output scores typically become prohibitively large.

\begin{figure}[t]
\centering
{\includegraphics[clip,width=.99\columnwidth]{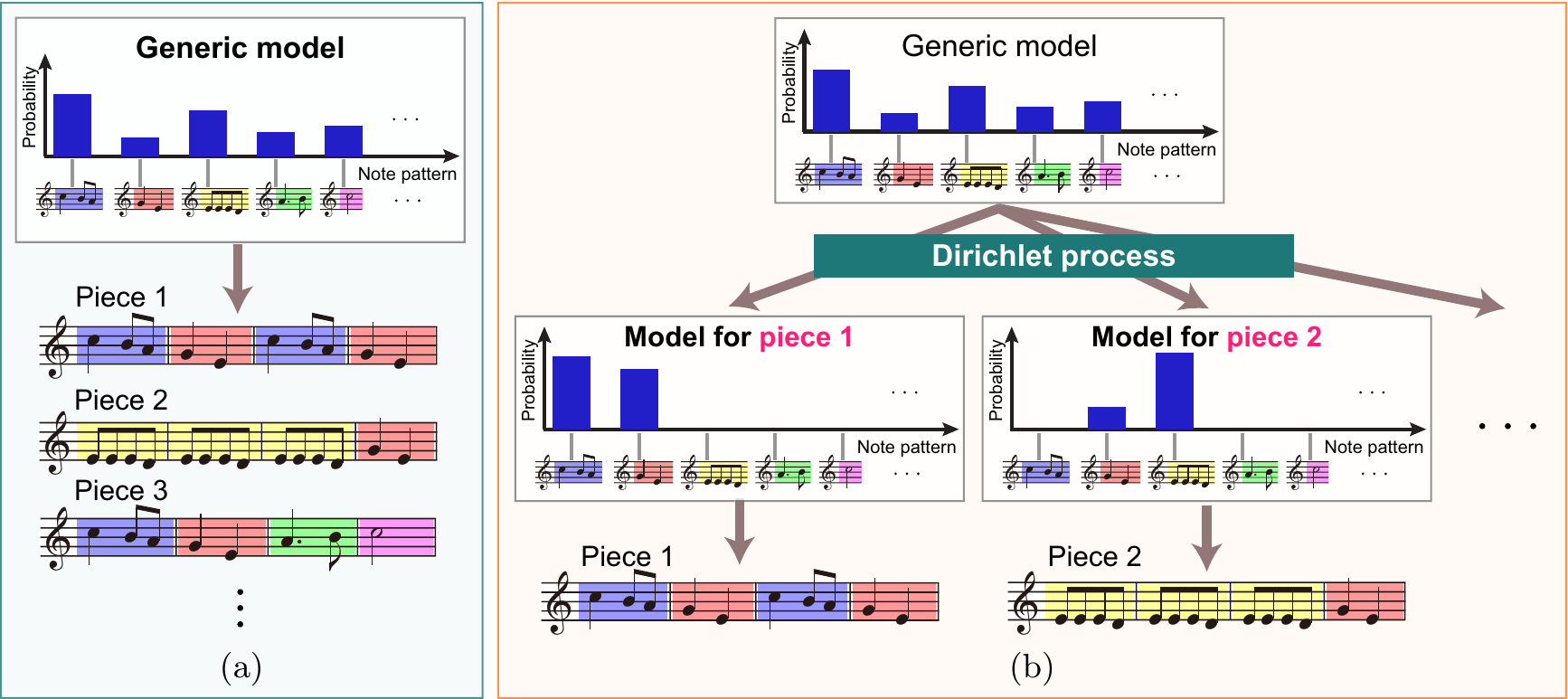}}
\vspace{-2mm}
\caption{Schematic illustrations of (a) a conventional score model and (b) the studied score model. In (a), all target scores are generated from a ``generic'' model. In (b), a piece-specific score model is generated for each individual score, where a repetitive structure is induced by a sparse distribution.}
\label{fig:Overview}
\vspace{-4mm}
\end{figure}
Recently, Nakamura et al.~\cite{Nakamura2016} proposed a statistical score model incorporating a repetitive structure and derived a tractable algorithm for music transcription based on the model.
A prominent feature of this method is that {\it piece-specific} score models for individual musical pieces are constructed and subsequently used for transcription, in contrast to the use of a {\it generic} score model for all target musical pieces in the conventional methods.
The model is formulated as a Bayesian model whereby a piece-specific score model is first generated from a generic score model and a musical score is then generated from the piece-specific model (Fig.~\ref{fig:Overview}).
This model is similar to the topic model of natural language based on the latent Dirichlet allocation (LDA) \cite{Blei2003}, where a word distribution of individual text is generated from a generic word distribution of a collection of text.
Instead of a word distribution, a distribution of note patterns (i.e.\ subsequences of notes corresponding to bars) is considered as a score model for generating musical scores.
The process for generating piece-specific models is described as a Dirichlet process \cite{Jordan2005} with a small concentration parameter, which induces sparse distributions of note patterns.
As a consequence of the sparseness of the piece-specific score model, repetitions of note patterns are naturally induced in the generated score.
It is important that the piece-specific score model is stochastically generated and thus the total model can describe unseen scores with an unfixed repetitive structure.
Moreover, by combining a process of modifying the generated note patterns, the model can also describe approximate repetitions, which are commonly seen in music practice.

In \cite{Nakamura2016}, the efficacy of the score model was tested in the task of the rhythm transcription of monophonic music.
It was found that both the Bayesian construction of score models and the note modification process improved the transcription accuracy, showing the potential of the approach.
It was also found that the computational costs were too large for practical applications.

The purpose of this study is to find a score model capturing repetitions that can be practically used for rhythm transcription.
We construct wider classes of Bayesian score models and conduct a systematic comparative evaluation of these models to find an optimal model in terms of transcription accuracy and computational costs.
While repetitions were considered in units of note patterns in \cite{Nakamura2016}, it is theoretically possible to consider repetitions in units of notes.
We construct Bayesian score models based on the note value Markov model (MM) \cite{Takeda2002} and the metrical MM \cite{Hamanaka2003,Raphael2002}, which are advantageous for computational efficiency compared to the note pattern MM considered in \cite{Nakamura2016}.
We apply the constructed score models to rhythm transcription and conduct evaluations to examine the effect of capturing repetitions and to reveal the best model for the task.

After introducing basic score models for rhythms (note value MM, metrical MM, and note pattern MM) and presenting statistical analyses on the repetitive structure in Sec.~\ref{sec:Preliminaries}, we formulate our models in Sec.~\ref{sec:Model}.
Bayesian extensions of the three types of Markov models and the integration of note modification processes (note divisions and onset shifts) are formulated there.
A rhythm transcription method based on the score models is presented in Sec.~\ref{sec:TranscriptionModels}.
Evaluations are carried out in Sec.~\ref{sec:Evaluation}, where models are compared in terms of the transcription error rate and computation time.
We also examine the influence of the parameterization of the relevant hyperparameters.
The data and source code used in this study are available at \url{https://bayesianscoremodel.github.io}.

%%%%%%%%%%%%%%%%%%%%%%%%%%%%%%%%%%%%%%%%%%%%%%
\section{Repetitive Structure and Sparseness of Piece-Specific Score Models}
\label{sec:Preliminaries}
%%%%%%%%%%%%%%%%%%%%%%%%%%%%%%%%%%%%%%%%%%%%%%

We here define the form of music data we address, introduce basic score models, and present statistical analyses that indicate the necessity of considering piece-specific score models in constructing musical score models capturing the repetitive structure of music.

%%%
\subsection{Musical Score Data}
\label{sec:Data}
%%%

%
\begin{table}[t]
\centering
\caption{Musical score data used in this study.}
\vspace{2mm}
{\tabcolsep = 3pt
\begin{tabular}{lccccccc}
\toprule
Dataset  & Original data & Pieces  & Bars & Notes \\
\midrule
Train    & Beatles            & $180$ & $16,746$ & $40,913$ \\
         & RWC                & $89$  & $11,449$ & $31,998$ \\
         & Contemporary J-pop & $132$ & $21,152$ & $66,073$ \\
         & Total              & $401$ & $49,347$ & $138,583$ \\
\midrule
Test     & Beatles            & $10$ & $924$   & $2,124$ \\
         & RWC                & $10$ & $1,452$ & $3,950$ \\
         & Contemporary J-pop & $10$ & $1,518$ & $4,672$ \\
         & Total              & $30$ & $3,894$ & $10,716$ \\
\bottomrule
\end{tabular}
}
\label{tab:Data}
%\vspace{-2mm}
\end{table}

The score data used in this study are extracted from a collection of vocal melodies of popular music.
Specifically, we use 99 pieces from the RWC popular music dataset \cite{RWC}, 190 pieces by The Beatles, and 142 contemporary Japanese popular music (J-pop) pieces.
To enable comparisons between different models, only pieces in 4/4 time are chosen, only rhythms (note onset positions) are used, the 16th-note length is used as the minimal beat resolution, and each piece is segmented into a half-note length.
In this step, we disregard rests, note onsets in finer beat resolutions, and segments without any note onsets.
In addition, for simplicity, if a pair of consecutive note onsets in two segments is more than a half-note length apart, a new onset is added at the beginning of the latter segment so that the maximum note length will be a half-note length.
Hereafter the obtained segments are called {\it bars}, that is, pieces are represented in 2/4 time.
The relative positions of note onsets in each bar in units of 16th notes are called the {\it metrical positions}, and the number of metrical positions $N_b$ ($=8$) is called the {\it bar length}.
We should remark that the time signature, bar length, and the manipulations made in preparing our data were chosen to balance the precision of the representation and the computational efficiency of numerical experiments; the extension for including longer durations and triplet notes is theoretically possible, as discussed in Sec.~\ref{sec:Discussion}.
Finally, we randomly split the data into two sets, one set for training the models and the other set for evaluating them, whose contents are shown in Table \ref{tab:Data}.

\begin{figure}[t]
\centering
{\includegraphics[clip,width=1.\columnwidth]{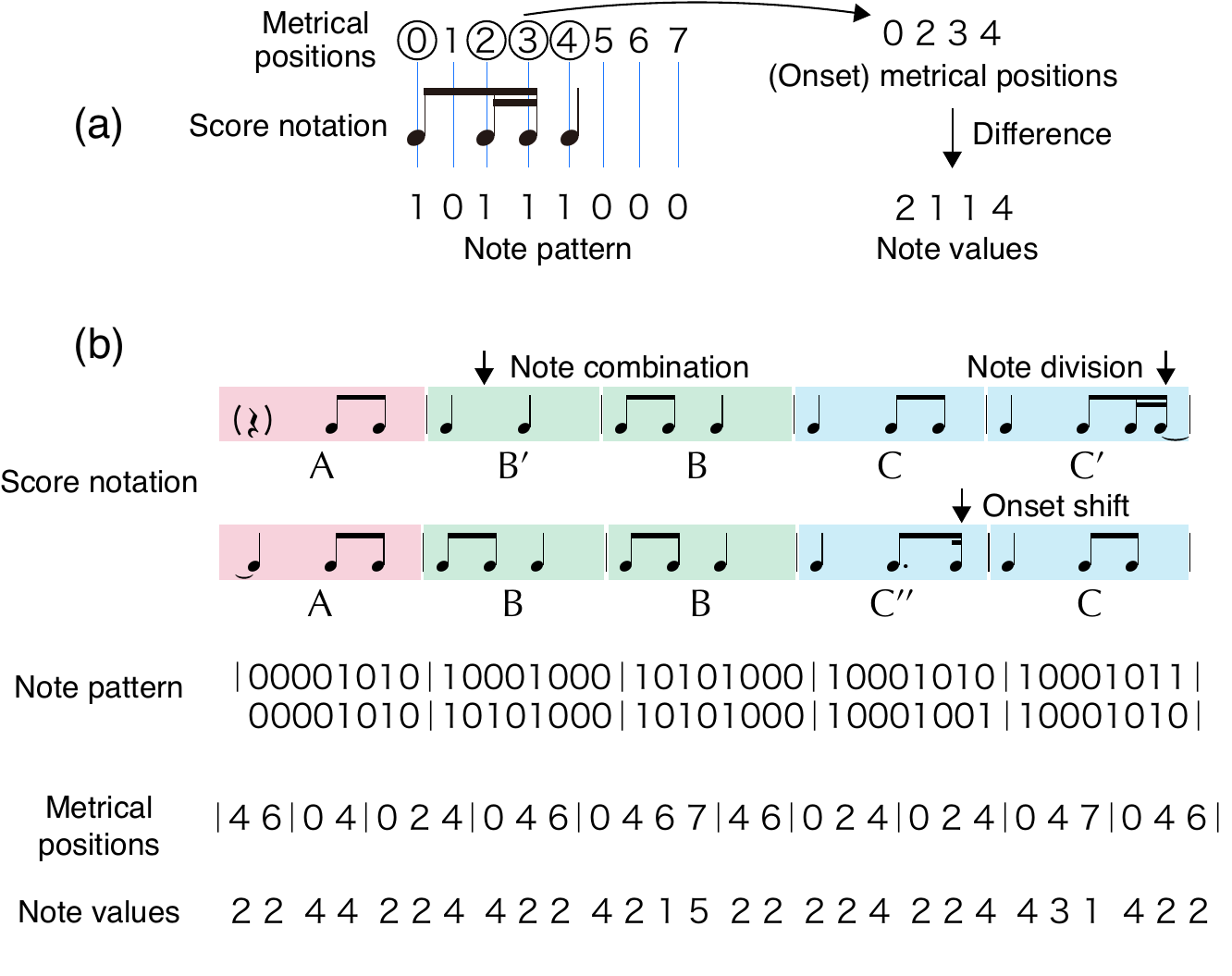}}
%\vspace{-7mm}
\caption{Representations of musical rhythms. (a) A one-bar example. (b) A longer example. The score is taken from ``Yasashii uta (tender song)'' by a J-pop band Mr.~Children.}
\label{fig:RhythmRepresentation}
%\vspace{-3mm}
\end{figure}
A {\it note pattern} (i.e.\ rhythmic pattern in a bar) can be represented as an $N_b$-bit binary vector $\sigma_1\sigma_2\cdots\sigma_{N_b}$, where $\sigma_b=1$ indicates an onset at metrical position $b$ and otherwise $\sigma_b=0$.
The onset time of a note measured from the beginning of a piece in units of 16th notes is called the {\it onset score time}.
A score-notated note length (difference in the onset score time) is called a {\it note value}.
As shown in Fig.~\ref{fig:RhythmRepresentation}, we can use one of three quantities (note value, metrical position, and note pattern) to represent musical rhythms.
These three representations are equivalent; however, the metrical position of the first note onset is lost in the note value representation.

We represent a musical score as a sequence of onset score times $(\tau_n)_{n=0}^{N}$ where $n$ represents the note onsets and $N$ is the number of notes ($N+1$ note onsets are necessary to define the lengths of $N$ notes).
The onset score times are described in units of 16th-note lengths.
In the example of Fig.~\ref{fig:RhythmRepresentation}(b), $\tau_0=4$, $\tau_1=6$, $\tau_2=8$, $\tau_3=12$, etc.
Their absolute values are unimportant but we assume that $\tau_n$ (mod $N_b$) represents the metrical position $b_n\in\{0,\ldots,N_b-1\}$ of the $n$\,th note onset.
The note value $r_n$ of the $n$\,th note ($n=1,\ldots,N$) is defined as $r_n=\tau_{n}-\tau_{n-1}$.
In the example of Fig.~\ref{fig:RhythmRepresentation}(b), $b_0=4$, $b_1=6$, $b_2=0$, $b_3=4$, etc.\ and $r_1=2$, $r_2=2$, $r_3=4$, etc.
Since the maximum note value and $N_b$ are $8$, we can also write $r_n=b_{n}-b_{n-1}$ if $b_{n}>b_{n-1}$ and $r_n=b_{n}-b_{n-1}+N_b$ otherwise.
Hereafter, we use the notation $\tau_{n_1:n_2}=(\tau_n)_{n=n_1}^{n_2}$ to indicate a sequence of variables.

%%%
\subsection{Basic Score Models}
\label{sec:BasicScoreModels}
%%%

A score model is a probabilistic generative model for sequences of onset score times that describe the probability $P(\tau_{0:N})$, or, equivalently, $P(b_{0:N})$ or $P(r_{1:N})$.
Here we review three basic score models: the note value MM, the metrical MM, and the note pattern MM.

\subsubsection{Note Value Markov Models (NoteMM)}
\label{sec:NoteMM}

The basic first-order note value MM (NoteMM1) \cite{Takeda2002} is defined with the initial and transition probabilities of note values:
\begin{equation}
P(r_1=r)=\pi_{{\rm ini}\,r},\quad P(r_n=r\,|\,r_{n-1}=r')=\pi_{r'r}.
\end{equation}
The zeroth-order note value MM (NoteMM0) is defined by using the unigram probabilities $P(r_n)$ instead of the transition probabilities $P(r_n|r_{n-1})$.
The second-order note value MM (NoteMM2) and higher-order models can be constructed by considering transition probabilities of the form $P(r_n|\,r_{n-k},\ldots,r_{n-1})$.
Since similar methods can be applied for lower-order and higher-order models \cite{Mari1997}, we describe only the first-order models in what follows.

\subsubsection{Metrical Markov Models (MetMM)}
\label{sec:MetMM}

The basic first-order metrical MM (MetMM1) \cite{Hamanaka2003,Raphael2002} generates a sequence of metrical positions as
\begin{equation}
P(b_0=b)=\chi_{{\rm ini}\,b},\quad P(b_n=b\,|\,b_{n-1}=b')=\chi_{b'b}.
\end{equation}
The onset score times $(\tau_n)_{n=0}^N$ and the note values $(r_n)_{n=1}^N$ are then given as
\begin{align}
&\tau_0=b_0,\quad
r_n=\tau_{n}-\tau_{n-1}=[b_{n-1},b_n],
\label{eq:BeatToScoreTime1}
\\
&[b_{n-1},b_n]:=
\begin{cases}
b_{n}-b_{n-1},&b_{n-1}<b_{n};\\
b_{n}-b_{n-1}+N_b,&b_{n-1}\geq b_{n}.
\end{cases}
\label{eq:BeatToScoreTime2}
\end{align}
\subsubsection{Note Pattern Markov Models (PatMM)}

The note pattern MMs can be described as a generalization of the metrical MMs.
We denote a note pattern of metrical positions as
\begin{equation}
B_k=(B_{k1},\ldots,B_{kI_k}),
\end{equation}
where $k$ is an index in the set of note patterns, $I_k$ denotes the number of note onsets in note pattern $B_k$, and each $B_{ki}$ ($i\in\{1,\ldots,I_k\}$) indicates a metrical position ($B_{ki}<B_{k(i{+}1)}$).
For example, the first two note patterns in Fig.~\ref{fig:RhythmRepresentation}(b) are represented as $(4,6)$ and $(0,4)$.
Note that $B_{k1}$ needs not be $0$: if $B_{k1}\neq0$ it indicates that the first note of pattern $B_k$ is a tied note.
This is an extension of the note pattern model used in \cite{Nakamura2016}, which cannot describe tied notes.

The basic first-order note pattern MM (PatMM1) is described as a two-level hierarchical MM \cite{Fine1998} with a state space indexed by a pair $(k,i)$ of note-pattern index $k$ and an internal index $i\in\{1,\ldots,I_k\}$ of notes in each pattern $k$.
The upper-level model generates a sequence of note patterns as
\begin{equation}
\Gamma_{{\rm ini}\,k}=P(k_1=k),\quad \Gamma_{k'k}=P(k_{m}=k\,|\,k_{m-1}=k').
\end{equation}
The lower-level model generates internal positions as
\begin{align}
&\gamma^{k}_{{\rm ini}\,i}=P(i_1=i\,|k)=\delta_{i1},
\\
&\gamma^{k}_{i'i}=P(i_{\ell+1}=i\,|\,i_\ell=i',k)=\delta_{(i'+1)i},
\\
&\gamma^{k}_{i\,{\rm end}}=P(i_{I_k}=i\,|k)=\delta_{i I_k},
\end{align}
where $\delta$ denotes Kronecker's delta.
The initial and transition probabilities of the total model are given as
\begin{align}
&\psi_{{\rm ini}(ki)}=P(k_1=k,i_1=i)=\Gamma_{{\rm ini}\,k}\gamma^{k}_{{\rm ini}\,i}=\Gamma_{{\rm ini}\,k}\delta_{i1},
\\
&\psi_{(k'i')(ki)}=P(k_{n}=k,i_{n}=i|k_{n-1}=k',i_{n-1}=i')
\notag\\
&~=\delta_{kk'}\gamma^{k}_{i'i}+\gamma^{k'}_{i'\,{\rm end}}\Gamma_{k'k}\gamma^{k}_{{\rm ini}\,i}
=\delta_{kk'}\delta_{i(i'+1)}+\delta_{i'I_{k'}}\Gamma_{k'k}\delta_{i1},
\end{align}
in which case we have $b_n=B_{k_ni_n}$.
The onset score times and note values are given as in Eqs.~(\ref{eq:BeatToScoreTime1}) and (\ref{eq:BeatToScoreTime2}).

%%%
\subsection{Repetitions and Sparseness}
\label{sec:RepetitionAndSparseness}
%%%

%
\begin{figure}[t]
\centering
{\includegraphics[clip,width=1\columnwidth]{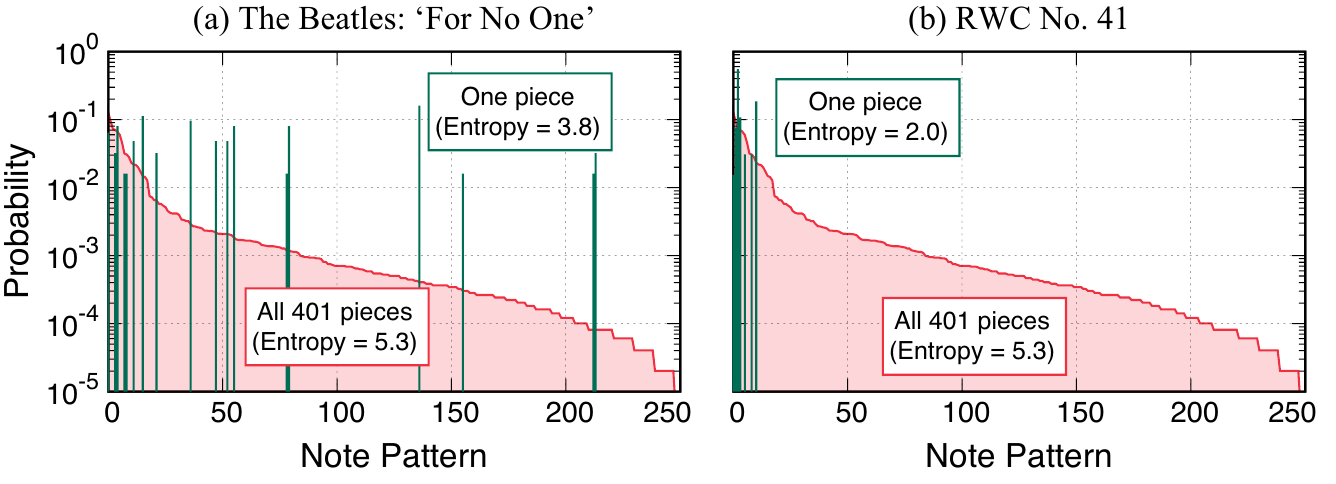}}
%\vspace{-5mm}
\caption{A generic distribution of note patterns obtained from the training data (background, red) and piece-specific distributions obtained from individual pieces in the same data (foreground, green). The note patterns are ordered according to the probability.}
\label{fig:PieceSpecificModel}
%\vspace{-3mm}
\end{figure}
\begin{figure}[t]
\centering
{\includegraphics[clip,width=.7\columnwidth]{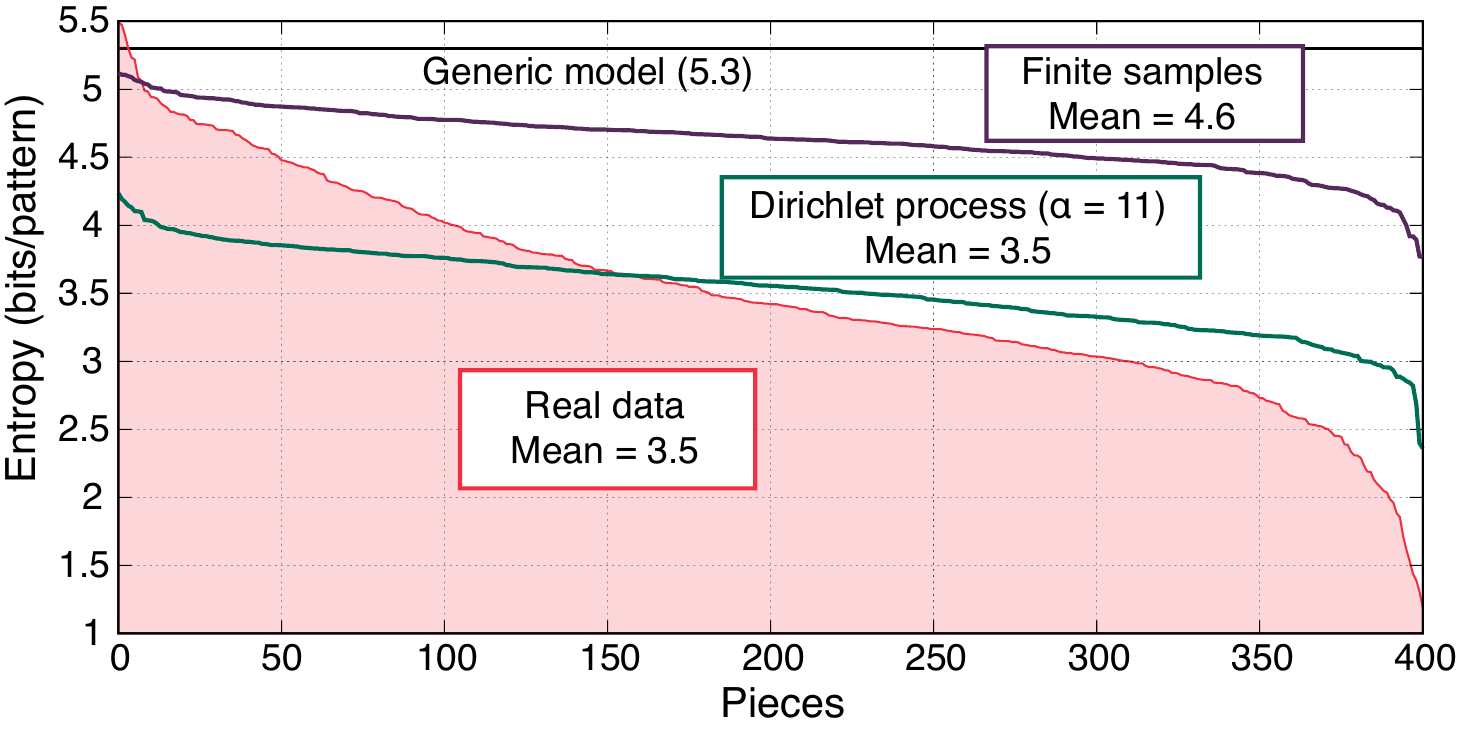}}
%\vspace{-2mm}
\caption{Distribution of the entropies of piece-specific distributions of note patterns, that of sequences sampled from the generic note pattern distribution, and that of distributions generated from the Dirichlet process. The pieces are ordered according to their entropy.}
\label{fig:EntropyDistributionPat0}
%\vspace{-3mm}
\end{figure}

To quantitatively see how a repetitive structure is reflected in the sparseness of piece-specific score models, a distribution ({\it generic} model) of note patterns obtained from all the pieces in our training dataset is shown with distributions ({\it piece-specific} models) obtained from individual pieces in Fig.~\ref{fig:PieceSpecificModel}.
We use only the training dataset for the analysis in this subsection.
The sparseness of the piece-specific distributions, which can be measured by the (Shannon) entropy, is evident.
From the entropies of the piece-specific distributions of all the pieces in Fig.~\ref{fig:EntropyDistributionPat0}, we see that most pieces have a much lower entropy compared to the generic distribution with an entropy of $5.3$.

The sparseness of the piece-specific distributions cannot be explained merely by using the statistical fluctuation of finite data.
To confirm this, we drew 401 sets of samples (note patterns) from the generic distribution, each of which corresponds to a piece in the data and has the same number of bars (the average number of bars was $123$).
The entropies calculated from the distributions obtained from the sampled data are also plotted in Fig.~\ref{fig:EntropyDistributionPat0}.
The result in Fig.~\ref{fig:EntropyDistributionPat0} shows that these distributions still have much higher entropies than the real data, indicating that the sparseness of the real piece-specific distributions originates from the repetitive structure.

\begin{figure}[t]
\centering
{\includegraphics[clip,width=.7\columnwidth]{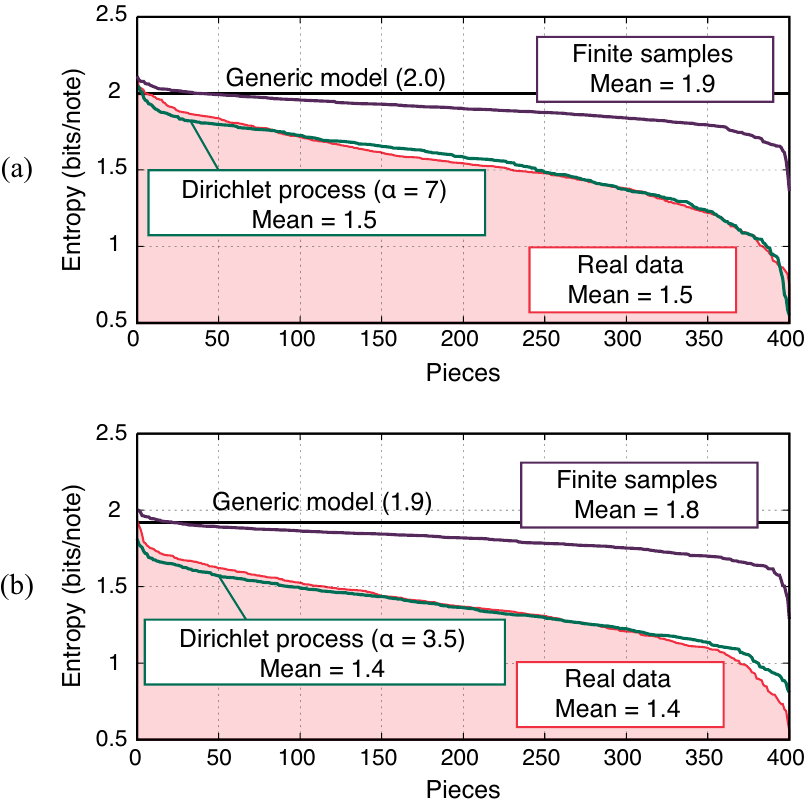}}
%\vspace{-2mm}
\caption{Distributions of the entropies of the piece-specific transition probabilities of (a) note values and (b) metrical positions. The corresponding distributions obtained from sampled data from the generic transition probabilities and those obtained from the transition probabilities generated from the Dirichlet process are also shown. The pieces are ordered according to their entropy.}
\label{fig:EntropyDistribution_Other}
%\vspace{-3mm}
\end{figure}
As mentioned above, rhythmic patterns can also be described as sequences of metrical positions or note values (Fig.~\ref{fig:RhythmRepresentation}) and repetitions at the note pattern level induce repetitions at the note level and vice versa.
This indicates that a repetitive structure is reflected in the sparseness of piece-specific note value MMs and metrical MMs, whose main parameters are the transition probabilities of note values and those of metrical positions, respectively.
In Figs.~\ref{fig:EntropyDistribution_Other}(a) and \ref{fig:EntropyDistribution_Other}(b), we plot the distributions of the entropies for these models: one for the real data and one for the sampled data from the generic model, respectively.
We can again observe that the entropies of the actual piece-specific transition probabilities are much lower than the entropy of the generic model and that this cannot be explained merely by the finite sample effect.

These results support our assertion that a repetitive structure in music can be useful for transcription from the viewpoint of generative modelling.
Lower entropy means a higher predictive ability in generative modelling.
Thus, if we can infer the piece-specific score model appropriately, the transcription will be easier due to the model's high predictive ability.

%%%
\subsection{Note Modifications for Approximate Repetitions}
\label{sec:AnalysisModification}
%%%

In music, repetitions may appear with modifications, leading to approximate repetitions.
The example in Fig.~\ref{fig:RhythmRepresentation}(b) illustrates the two most common types of modifications: note division/combination and onset shift \cite{Nakamura2016,Siorosa2018}.
In a note division/combination (or simply division/combination), a note onset is inserted/deleted while the other note onsets are unchanged.
These modifications often appear in sung melodies where the number of syllables varies in repeated phrases or sections.
In an onset shift (or simply shift), a note onset is shifted forward or backward in time.
This is another typical type of rhythmic variation that includes syncopations.

Modelling these modifications for describing musical scores with approximate repetitions can be useful for transcription because by identifying modifications we can induce more repetitions in data, which can lead to a higher predictive ability.
The concept of approximate repetitions is also important for humans to understand or transcribe music \cite{SweetAnticipation}.
The example in Fig.~\ref{fig:RhythmRepresentation}(b) cannot be recognized as repeated phrases if modified note patterns are considered to be completely different ones.

%%%%%%%%%%%%%%%%%%%%%%%%%%%%%%%%%%%%%%%%%%%%%%
\section{Proposed Score Models}
\label{sec:Model}
%%%%%%%%%%%%%%%%%%%%%%%%%%%%%%%%%%%%%%%%%%%%%%

%%%
\subsection{Overview of the Studied Models}
%%%

%
\begin{table}[t]
\centering
\caption{List of the constructed score models based on the note value Markov model (MM). In the acronyms, the number indicates the order of the Markov model, `S' and `D' indicate note shift and note division, and `B' indicates Bayesian extension.}
\vspace{2mm}
{\tabcolsep = 3pt
\begin{tabular}{lcccccc}\toprule
Acronym    & MM order & Bayesian     & Modification\\
\midrule
NoteMM0    & 0th      &              & --- \\
NoteMM0B   & 0th      & $\checkmark$ & --- \\
NoteMM1    & 1st      &              & --- \\
NoteMM1B   & 1st      & $\checkmark$ & --- \\
NoteMM1SB  & 1st      & $\checkmark$ & Shift only \\
NoteMM1DB  & 1st      & $\checkmark$ & Division only \\
NoteMM1SDB & 1st      & $\checkmark$ & Shift \& division \\
NoteMM2    & 2nd      &              & --- \\
NoteMM2B   & 2nd      & $\checkmark$ & --- \\
\bottomrule
\end{tabular}
}
\label{tab:ModelList}
%\vspace{-2mm}
\end{table}
Here, we explain the proposed score models.
We construct score models based on three basic models: the note value MM, the metrical MM, and the note pattern MM.
Each basic model is extended in a Bayesian manner to describe the repetitive structure and combined with note modification models to describe approximate repetitions.
The incorporation of note modifications is described in Sec.~\ref{sec:NoteModificationModel} and the Bayesian extensions are formulated in Sec.~\ref{sec:BayesianExtension}.
The list of constructed models based on the note value MM and their acronyms are summarized in Table \ref{tab:ModelList}.
Similarly we construct 9 models (MetMM0, MetMM0B, $\ldots$, MetMM2B) based on the metrical MM and 7 models (PatMM0, PatMM0B, $\ldots$, PatMM1SDB) based on the note pattern MM.
We do not consider the second-order note pattern MM and its extensions due to their impractical computational cost.

%%%
\subsection{Incorporation of Note Modifications}
\label{sec:NoteModificationModel}
%%%

%
\subsubsection{Note Modification Operations}
\label{sec:ModificationOperations}
\begin{figure}[t]
\centering
{\includegraphics[clip,width=.6\columnwidth]{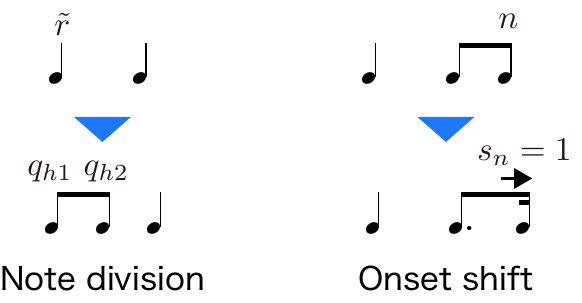}}
%\vspace{-2mm}
\caption{The note division operation (left) and the onset shift operation (right).}
\label{fig:Modification}
\vspace{-3mm}
\end{figure}
We consider the most common types of note modifications, onset shifts, note divisions, and note combinations (Sec.~\ref{sec:AnalysisModification}).
As note divisions and combinations are the opposite of each other, we only consider note divisions in our models.
In \cite{Nakamura2016}, only onset shifts for notes on bar onsets (i.e.\ syncopations) were considered.
Here, we generalize the idea and consider onset shifts for any notes.

We denote the shift of the onset score time of the $n$\,th note as $s_n$ and the note value and score time of the $n$\,th note before the application of onset shifts are notated as $\tilde{r}_n$ and $\tilde{\tau}_n$, respectively (Fig.~\ref{fig:Modification} (right)).
An onset shift is then described as
\begin{equation}
\tilde{\tau}_n\to\tau_n=\tilde{\tau}_n+s_n\quad{\rm or}\quad
\tilde{r}_n\to r_n=\tilde{r}_n+s_n-s_{n-1}.
\end{equation}
The shift variable $s_n$ describes the amount of the shift and takes values in the range $[-N_b+1,N_b-1]$ with probability
\begin{equation}
\xi_s=P(s_n=s).
\end{equation}
We assume a constraint $-\tilde{r}_n<s_n\leq \tilde{r}_n$ to make the onset shift interpretable as a modification.
Musically, we should also have $r_n=\tilde{r}_n+s_n-s_{n-1}>0$ for all $n$.
%For mathematical simplicity, we do not impose a constraint to assure this, but the observed data always satisfy this.

In the most general case, a note division is described as an operation $\tilde{r}\to\bm q_h$ that maps a note value $\tilde{r}$ to a sequence of note values $\bm q_h=q_{h1}\cdots q_{hG_h}$ (Fig.~\ref{fig:Modification} (left)).
Here, $h$ labels the patterns of note divisions and $G_h$ is the number of notes after a division.
The division probability can be written as
\begin{equation}
\zeta_{\tilde{r}h}=P(\tilde{r}\to \bm q_h).
\end{equation}
We must have $\zeta_{\tilde{r}h}=0$ unless $\tilde{r}=q_{h1}+\cdots+q_{hG_h}$.
Using an index $g\in\{1,\ldots,G_h\}$, the pair of indices $(h,g)$ can specify the note value after applying the note division as $r_n=q_{h_ng_n}$.

When we combine onset shifts and note divisions, we first apply note divisions and then apply an onset shift to each of the divided notes.
The resulting notes are indexed by $(\tilde{r},h,g,s)$, and their note values are given as
\begin{equation}
r_n=q_{h_ng_n}+s_n-s_{n-1}.
\end{equation}
In this study, we only consider note divisions into two notes, that is, $G_h\leq2$, to avoid impractical computational costs.

\subsubsection{Score Models with Note Modifications}
\label{sec:ScoreModelsWithNoteModifications}

Note value MMs can be extended to incorporate note modifications as follows.
The model incorporating onset shifts (NoteMM1S) is described by an HMM with a state space indexed by $z_0=s_0$ and $z_n=(\tilde{r}_n,s_n)$ ($n=1,\ldots,N$), where $\tilde{r}_n$ indicates the note value before modification and $s_n$ indicates the amount of the shift.
The initial and transition probabilities are given as
\begin{align}
&P(z_0=s)=\xi_s,\quad P(z_1=(\tilde{r},s))=\pi_{{\rm ini}\,\tilde{r}}\,\xi_s,
\\
&P(z_{n}=(\tilde{r},s)\,|\,z_{n-1}=(\tilde{r}',s'))=\pi_{\tilde{r}'\tilde{r}}\,\xi_s,
\end{align}
and the final output is obtained from the output probability
\begin{equation}
P(r_n|z_{n-1},z_n)=\mathbb{I}(r_n=\tilde{r}_n+s_n-s_{n-1}),
\end{equation}
where $\mathbb{I}({\cal C})$ is $1$ when expression ${\cal C}$ is true and is $0$ otherwise.

The note value MM incorporating note divisions (NoteMM1D) is described by an HMM with a state space indexed by $z=(\tilde{r},h,g)$, where $\tilde{r}$ indicates the note value before note division and $h$ and $g$ indicate a note-division pattern and its internal note position, respectively (see Sec.~\ref{sec:ModificationOperations}).
The initial and transition probabilities are given as
\begin{align}
&P(z_1=(\tilde{r},h,g))=\pi_{{\rm ini}\,\tilde{r}}\zeta_{\tilde{r}h}\delta_{g1},
\\
&P(z_{n}=(\tilde{r},h,g)\,|\,z_{n-1}=(\tilde{r}',h',g'))
\notag\\
&\quad=\delta_{\tilde{r}\tilde{r}'}\delta_{hh'}\delta_{g(g'+1)}+\delta_{g'G_{h'}}\pi_{\tilde{r}'\tilde{r}}\zeta_{\tilde{r}h}\delta_{g1},
\end{align}
and the final output is obtained from the output probability
\begin{equation}
P(r_n|z_n)=\mathbb{I}(r_n=q_{h_ng_n}).
\end{equation}

The note value MM incorporating both onset shifts and note divisions (NoteMM1SD) is obtained by combining the above two models.
It is described by an HMM with a state space indexed by $z_0=s_0$ and $z_n=(\tilde{r}_n,h_n,g_n,s_n)$ ($n=1,\ldots,N$) and the initial and transition probabilities are given as
\begin{align}
&P(z_0=s)=\xi_s,~~
P(z_1=(\tilde{r},h,g,s))=\pi_{{\rm ini}\,\tilde{r}}\zeta_{\tilde{r}h}\delta_{g1}\xi_s,
\notag
\\
&P(z_{n}=(\tilde{r},h,g,s)\,|\,z_{n-1}=(\tilde{r}',h',g',s'))
\notag\\
&\quad=\big(\delta_{\tilde{r}\tilde{r}'}\delta_{hh'}\delta_{g(g'+1)}+\delta_{g'G_{h'}}\pi_{\tilde{r}'\tilde{r}}\zeta_{\tilde{r}h}\delta_{g1}\big)\xi_s.
\end{align}
The final output is obtained from the output probability
\begin{equation}
P(r_n|\,z_{n-1},z_n)=\mathbb{I}(r_n=q_{h_ng_n}+s_n-s_{n-1}).
\end{equation}

The model parameters of the (non-Bayesian) first-order note value models are summarized as follows.
Hereafter we use $r$ instead of $\tilde{r}$ to unify the notation.
Let us write $\bm\pi_{\rm ini}=(\pi_{{\rm ini}\,r})_r$, $\bm\pi_r=(\pi_{rr'})_{r'}$, $\bm\zeta_{r}=(\zeta_{rh})_h$, and $\bm\xi=(\xi_s)_s$.
\begin{itemize}
\item NoteMM1: $\bm\pi_{\rm ini}$, $(\bm\pi_r)_r$.
\item NoteMM1S: $\bm\pi_{\rm ini}$, $(\bm\pi_r)_r$, $\bm\xi$.
\item NoteMM1D: $\bm\pi_{\rm ini}$, $(\bm\pi_r)_r$, $(\bm\zeta_{r})_{r}$.
\item NoteMM1SD: $\bm\pi_{\rm ini}$, $(\bm\pi_r)_r$, $(\bm\zeta_{r})_{r}$, $\bm\xi$.
\end{itemize}
We can similarly formulate four models (MetMM1, MetMM1S, MetMM1D, and MetMM1SD) based on the metrical MM with or without the note modification process and four models (PatMM1, PatMM1S, PatMM1D, and PatMM1SD) based on the note pattern MM.
See the Supplemental Material for the explicit formulation.

%%%
\subsection{Bayesian Extensions Based on Dirichlet Processes}
\label{sec:BayesianExtension}
%%%

We extend the score models in a Bayesian manner to formulate the generative process of piece-specific models from a generic model.
We assume that piece-specific models and generic models have the same architecture but different parameterizations.
For NoteMM1SD, the generative process can be formulated by assigning prior distributions to the model parameters:
\begin{align}
\bm\pi_{\rm ini}&\sim{\rm DP}(\alpha_{\rm ini},\bar{\bm\pi}_{\rm ini}),\quad
\bm\pi_{r}\sim{\rm DP}(\alpha_{\pi},\bar{\bm\pi}_{r}),
\notag\\
\bm\zeta_{r}&\sim{\rm DP}(\alpha_{\zeta},\bar{\bm\zeta}_{r}),\quad
\bm\xi\sim{\rm DP}(\alpha_{\xi},\bar{\bm\xi}),
\end{align}
where DP denotes a Dirichlet process \cite{Jordan2005}; $\bar{\bm\pi}_{\rm ini}$, $\bar{\bm\pi}_{r}$, etc.\ denote base distributions; and $\alpha_{\rm ini}$, $\alpha_\pi$, etc.\ denote concentration parameters.
Since the distributions are finite discrete distributions, the Dirichlet process can be described as Dirichlet distributions.
For example,
\begin{equation}
\bm\pi_{r}\sim{\rm DP}(\alpha_{\pi},\bar{\bm\pi}_{r})={\rm Dir}(\alpha_{\pi}\bar{\bm\pi}_{r}).
\end{equation}
The expectation value of the distributions generated by a Dirichlet process is equal to the base distribution \cite{Jordan2005}.
Based on this fact, we can interpret the distributions ($\bm\pi_{r}$, $\bm\zeta_{r}$, etc.) generated by the Dirichlet process as the piece-specific models and the base distributions ($\bar{\bm\pi}_{r}$, $\bar{\bm\zeta}_{r}$, etc.) as the generic model representing the average of all the piece-specific models.
The distributions generated from a Dirichlet process become sparser as we use a smaller concentration parameter.
The parameters of the note modification models $\bm\zeta_{r}$ and $\bm\xi$ are also considered for individual pieces to describe the situation in which the note modifications used in each piece have different statistical tendencies.

We can similarly construct Bayesian extensions of metrical MMs and note pattern MMs.
For metrical MMs, we have
\begin{align}
\bm\chi_{\rm ini}&\sim{\rm DP}(\alpha_{\rm ini},\bar{\bm\chi}_{\rm ini}),\quad
\bm\chi_{b}\sim{\rm DP}(\alpha_{\chi},\bar{\bm\chi}_{b}),
\\
\bm\zeta_{r}&\sim{\rm DP}(\alpha_{\zeta},\bar{\bm\zeta}_{r}),\quad
\bm\xi\sim{\rm DP}(\alpha_{\xi},\bar{\bm\xi}),
\end{align}
and for note pattern MMs, we have
\begin{align}
\bm\Gamma_{\rm ini}&\sim{\rm DP}(\alpha_{\rm ini},\bar{\bm\Gamma}_{\rm ini}),\quad
\bm\Gamma_{k}\sim{\rm DP}(\alpha_{\Gamma},\bar{\bm\Gamma}_{k}),
\\
\bm\zeta_{r}&\sim{\rm DP}(\alpha_{\zeta},\bar{\bm\zeta}_{r}),\quad
\bm\xi\sim{\rm DP}(\alpha_{\xi},\bar{\bm\xi}).
\end{align}
where $\bar{\bm\chi}_{\rm ini}$, $\bar{\bm\chi}_{b}$, $\bar{\bm\Gamma}_{\rm ini}$, $\bar{\bm\Gamma}_{k}$, etc.\ denote the base distributions corresponding to the generic models and $\alpha_{\rm ini}$, $\alpha_\chi$, $\alpha_{\rm ini}$, $\alpha_\Gamma$, etc.\ denote the corresponding concentration parameters.

In Fig.~\ref{fig:EntropyDistributionPat0}, the entropies of the distributions of the note patterns (zeroth-order note pattern MMs) generated by a Dirichlet process are shown, where the base distribution is the generic model and the concentration parameter is adjusted to match the average entropy with the real data.
Although the obtained distribution of entropies does not very well match the actual distribution, these distributions overlap much more than the actual distribution does with the entropy distribution of the sampled data from the generic model.
Similarly, in Figs.~\ref{fig:EntropyDistribution_Other}(a) and \ref{fig:EntropyDistribution_Other}(b), we plot the distributions of the entropies for the piece-specific note value MMs and metrical MMs obtained from the Dirichlet processes.
We see that the Dirichlet processes can generate transition probabilities with a similar entropy to the entropies of the actual piece-specific transition probabilities.
These results indicate that we can use small concentration parameters to account for the sparse piece-specific transition probabilities in the real data.

%%%%%%%%%%%%%%%%%%%%%%%%%%%%%%%%%%%%%%%%%%%%%%
\section{Rhythm Transcription Method}
\label{sec:TranscriptionModels}
%%%%%%%%%%%%%%%%%%%%%%%%%%%%%%%%%%%%%%%%%%%%%%

The aim of a rhythm transcription method is to estimate the score onset times $\tau_{0:N}$ (or note values $r_{1:N}$) from input MIDI data with note onset times $t_{0:N}$ (or durations $d_{1:N}$, where $d_n=t_n-t_{n-1}$) that are represented in units of seconds.
We first construct a musical performance generative model that describes the probability $P(\tau_{0:N},t_{0:N})$ or $P(r_{1:N},d_{1:N})$ by combining a score model (one of the aforementioned models) and a performance model explained in Sec.~\ref{sec:PerformanceModel}.
We then derive a rhythm transcription algorithm that can estimate the score by maximizing the probability $P(\tau_{0:N}|t_{0:N})$ or $P(r_{1:N}|d_{1:N})$, which is explained in Sec.~\ref{sec:AlgorithmForRhythmTranscription}.
For the Bayesian score models, the model parameters for the piece-specific model are learned in the transcription step, as described in Sec.~\ref{sec:ParameterLearning}.

%%%
\subsection{Performance Model}
\label{sec:PerformanceModel}
%%%

A performance (timing) model describes the generative process of durations $d_{1:N}$ from note values $r_{1:N}$, which gives the probability $P(d_{1:N}|r_{1:N})$.
We consider a simplified model with a constant (inverse) tempo $v$.
Given a note value $r_n$ for the $n$\,th note, the corresponding duration $d_n$ is generated as
\begin{equation}
d_n\sim{\rm Gauss}(vr_n,\sigma_t^2),
\end{equation}
where ${\rm Gauss}(\mu,\Sigma)$ denotes a Gaussian distribution with mean $\mu$ and variance $\Sigma$, and $\sigma_t$ represents the amount of onset time deviations.
The onset times $t_{0:N}$ can be determined by the durations $d_{1:N}$ up to an initial time $t_0$, which is insignificant.

In a real music transcription situation, the global tempo is unknown and the tempo changes over time, especially in solo performances and classical musical performances.
The performance model can be generalized to describe these situations by introducing an additional process to generate a sequence of local tempos $v_n$ \cite{Nakamura2017}.
To focus on the effect of musical score models, we only consider a constant and known tempo in this study.

%%%
\subsection{Parameter Learning}
\label{sec:ParameterLearning}
%%%

For non-Bayesian score models, the model parameters are pretrained or preset and are fixed during the transcription step.
The initial and transition probabilities can be directly learned from musical score data.
The probabilities $\bar{\bm\zeta}_{r}$ and $\bar{\bm\xi}$ of the modification models cannot be directly learned from musical score data, and thus they are manually preset to reflect the belief that modifications are rare.
Specifically, we assign probabilities $\bar{\xi}_0$ and $\bar{\zeta}_0$, which are slightly less than unity, for cases without modifications, and other entries in $\bar{\bm\zeta}_{r}$ and $\bar{\bm\xi}$ are assumed to have uniform probabilities.

For Bayesian models, the hyperparameters of the prior distributions except for the concentration parameters are pretrained or preset to the parameterization of the corresponding non-Bayesian models.
The concentration parameters can in principle be optimized according to the transcription accuracy.
In our evaluation in Sec.~\ref{sec:EvaluationScoreModel}, they are preset to a fixed value, and how the values of the concentration parameters for the unigram or transition probabilities influence the transcription accuracy is studied in Sec.~\ref{sec:InfluenceHyperparameter}.

The other parameters of the Bayesian models, namely the parameters and internal variables of a piece-specific score model, are treated as variables and estimated in the transcription step according to the input MIDI data.
Let $X=d_{1:N}$ (or $t_{0:N}$) denote the input MIDI data, $Z$ denote the set of internal variables of a score model including note values, $\Theta$ denote the set of parameters of the piece-specific score model, and $\Xi$ denote the set of hyperparameters.
For NoteMM1S, for example, $Z=(s_0,\tilde{r}_1,s_1,\ldots,\tilde{r}_N,s_N)$, $\Theta=(\bm\pi_{\rm ini},\bm\pi_r,\bm\xi)$, and $\Xi=(\alpha_{\rm ini},\bar{\bm\pi}_{\rm ini},\alpha_\pi,\bar{\bm\pi}_r,\alpha_\xi,\bar{\bm\xi})$.
The generative process can be summarized as
\begin{equation}
P(X,Z,\Theta\,|\Xi)=P(X|Z)P(Z|\Theta)P(\Theta|\Xi).
\label{eq:CompleteGenerativeProcess}
\end{equation}
Here, the first factor on the right-hand side represents the performance model (Sec.~\ref{sec:PerformanceModel}), the second factor represents the (non-Bayesian part of the) score model (Secs.~\ref{sec:BasicScoreModels} and \ref{sec:NoteModificationModel}), and the last factor represents the prior distributions (Sec.~\ref{sec:BayesianExtension}).
Parameters $\Theta$ of the piece-specific score model reflecting both the piece-specific distribution of the rhythmic patterns inferred from the observation $X$ and the prior distribution, which induces the sparseness of $\Theta$, can be obtained by maximizing Eq.~(\ref{eq:CompleteGenerativeProcess}).
Since an analytical solution to this inference problem is impossible, as is typically the case for Bayesian models with latent variables, we use the Gibbs sampling method to estimate $\Theta$ \cite{Bishop}.
Using the Gibbs sampling method, we can draw samples from the posterior probability $P(Z,\Theta\,|\,X,\Xi)$.
After some iterations, we obtain the optimal parameters $\Theta$ that maximizes the likelihood $P(X|\Theta)=\sum_ZP(X|Z)P(Z|\Theta)$ among the sampled parameters.
For a simple case (the first-order metrical MM), an explicit algorithm using Gibbs sampling is presented in \ref{sec:ExplicitAlgorithm}.
Although the algorithmic details of the Gibbs sampling method are important for the implementation of the models, they are derived by means of standard techniques and are not essential for understanding the main results of this study.
Thus, for the other models, we present those details with the explicit formulations of the integrated models in the Supplemental Material.

%%%
\subsection{Algorithms for Rhythm Transcription}
\label{sec:AlgorithmForRhythmTranscription}
%%%

Our generative models for rhythm transcription constructed as combinations of a score model and a performance model are HMMs whose latent states are the variables of the score model and whose observed variables are onset times or durations of input (human-performed) MIDI data.
Thus, the standard Viterbi algorithm \cite{Rabiner1989} can be used to estimate the variables of the score model, including the note values or metrical positions, from the input MIDI data.
For Bayesian models, the model parameters are first estimated from the input signal using the method explained in Sec.~\ref{sec:ParameterLearning}, and the latent variables are then estimated by the Viterbi algorithm.
As an example, an explicit algorithm for rhythm transcription based on the first-order Bayesian metrical MM (MetMM1B) is presented in \ref{sec:ExplicitAlgorithm}.

For some of the present models, especially PatMM1DB and PatMM1SDB, the latent state space is huge, and the Viterbi algorithm requires an impractical amount of computation time.
To address this issue, we apply a beam search and only retain the top-$W$ most likely states at each Viterbi update ($W$ is a preset number called the beam width).
Using $N_Z$ to represent the number of latent states, the computational costs are reduced to $O(N_ZW)$ from $O(N_Z^2)$ via this technique, while the chance of obtaining the globally optimal solution decreases for smaller $W$s.
Since the inferential precision increases as we increase $W$, its value should be set to a practical upper limit for tractable computation.
For Bayesian extensions of these models, we also apply a similar method in the forward algorithm used for estimating the posterior probability.

%%%%%%%%%%%%%%%%%%%%%%%%%%%%%%%%%%%%%%%%%%%%%%
\section{Evaluation}
\label{sec:Evaluation}
%%%%%%%%%%%%%%%%%%%%%%%%%%%%%%%%%%%%%%%%%%%%%%

To evaluate the performance of the studied models, we conducted two evaluation experiments.
In the first experiment (Sec.~\ref{sec:EvaluationScoreModel}), in order to compare the effects of different model architectures, the predictive ability of the non-Bayesian score models is evaluated.
In the second experiment (Sec.~\ref{sec:EvaluationTranscription}), the transcription accuracies of the models are measured to examine the effects of the Bayesian extensions and modification models.
We also examine the influence of the hyperparameters of the studied models in Sec.~\ref{sec:InfluenceHyperparameter}.
Using the measurement of the computation time of the models, in Sec.~\ref{sec:Discussion}, we discuss which models are practically important.

%%%
\subsection{Setup}
\label{sec:Setup}
%%%

The initial and transition probabilities of all the models were trained with the same training dataset (Sec.~\ref{sec:Data}).
The probabilities were estimated by the maximum likelihood method with an additive smoothing constant of $0.1$.
We confirmed that the results were not sensitive to the smoothing constant, except for the transition probabilities of PatMM1 for which there were more parameters than the number of training samples.
For this model we applied a linear interpolation with the unigram probabilities ($0.8$ unigram probability $+$ $0.2$ transition probability) that was roughly optimized w.r.t.\ the transcription accuracy.

We used for the evaluations the test dataset consisting of melodies taken from 30 pieces of popular music (Sec.~\ref{sec:Data}).
For the transcription experiments, we collected performed MIDI data played by amateur musicians and recorded them with a digital piano.
Four musicians participated and were exclusively assigned some pieces in the test dataset.
The first 40 bars of each piece were used, and the musicians were asked to play the scores in a tempo of $105$ bpm (beats per minute).
For systematic examinations, we also used synthetic MIDI data generated by the performance timing model in Sec.~\ref{sec:PerformanceModel} with $\sigma_t=0.04$ sec and a tempo of $144$ bpm (the value of $\sigma_t$ was determined to simulate human performance and the tempo was determined so that the transcription difficulty roughly matches that for real MIDI data).
All pieces in the test dataset were used for the synthetic data.

%%%
\subsection{Evaluation of the Score Models}
\label{sec:EvaluationScoreModel}
%%%

\begin{table}[t]
\centering
\caption{Cross entropies (CEs) (bits/symbol) of the non-Bayesian score models.}
\vspace{2mm}
{\tabcolsep = 4pt
\begin{tabular}{lclclc}\toprule
Model      & CE     & Model      & CE     & Model      & CE     \\
\midrule
NoteMM0    & $2.29$ & NoteMM1    & $2.07$ & NoteMM2    & $1.93$ \\
MetMM0     & $2.63$ & MetMM1     & $2.01$ & MetMM2     & $1.87$ \\
PatMM0     & $1.97$ & PatMM1     & $1.89$ \\
\bottomrule
\end{tabular}
}
\label{tab:CrossEntropy}
%\vspace{-2mm}
\end{table}

We evaluated the non-Bayesian score models listed in Table \ref{tab:CrossEntropy} in terms of the cross entropy, which is a standard measure for quantifying the performance of language models.
Lower entropy means a higher predictive ability.
For all classes of models (note value MM, metrical MM, and note pattern MM) the cross entropy decreased as we increased the order.
Comparing the note value MMs and metrical MMs, the former outperformed when the order was zero, but the latter outperformed for the higher-order cases.
The zeroth-order note pattern MM has lower cross entropy than the first-order note value MM and metrical MM.
This demonstrates the strength of the note pattern models, which can capture long-range sequential dependence at the note level.
The reason the first-order note pattern MM was worse than the second-order metrical MM is likely because the data sparseness was more severe for the former model.

%%%
\subsection{Evaluation of the Transcription Accuracy}
\label{sec:EvaluationTranscription}
%%%

We used the synthetic data in addition to the real data of human performances to clearly examine the effects of the score models.
The following setups were used for the rhythm transcription algorithms based on the studied models.
For the evaluations using the synthetic data, we fixed the parameter $\sigma_t$ to its actual value ($0.04$ sec) used for the synthesis; and for the real data, we set $\sigma_t=0.035$ sec, which was roughly optimized in the preliminary experiments.
All the concentration parameters were set to $10$.
To estimate the model parameters of the Bayesian models, we iterated the Gibbs sampling $100$ times, except for the models with too large of computational costs.
For these models (MetMM1SDB, PatMM1DB, and PatMM1SDB), we iterated $20$ times.
The influence of varying the values for these parameters is investigated in Sec.~\ref{sec:InfluenceHyperparameter}.
The probability parameters for note modifications were set as $\bar{\xi}_0=\bar{\zeta}_0=0.9$, reflecting the assumption that note modifications are rare.
For PatMM1DB and PatMM1SDB, we set the beam width as $W=200$.
This value was a practical upper limit to run all experiments in a reasonable amount of time.
We use the error rate (i.e.\ the ratio of the number of notes with incorrectly estimated note values and the total number of notes) as an evaluation measure.
Since note values can be obtained from metrical positions, this evaluation measure can be applied universally for the transcribed results of all the models.
Since the transcription results depend on random numbers used for Gibbs sampling for the Bayesian models, we run the methods 10 times with different random number seeds and obtained the average error rates.

\begin{figure}[t]
\centering
{\includegraphics[clip,width=.98\columnwidth]{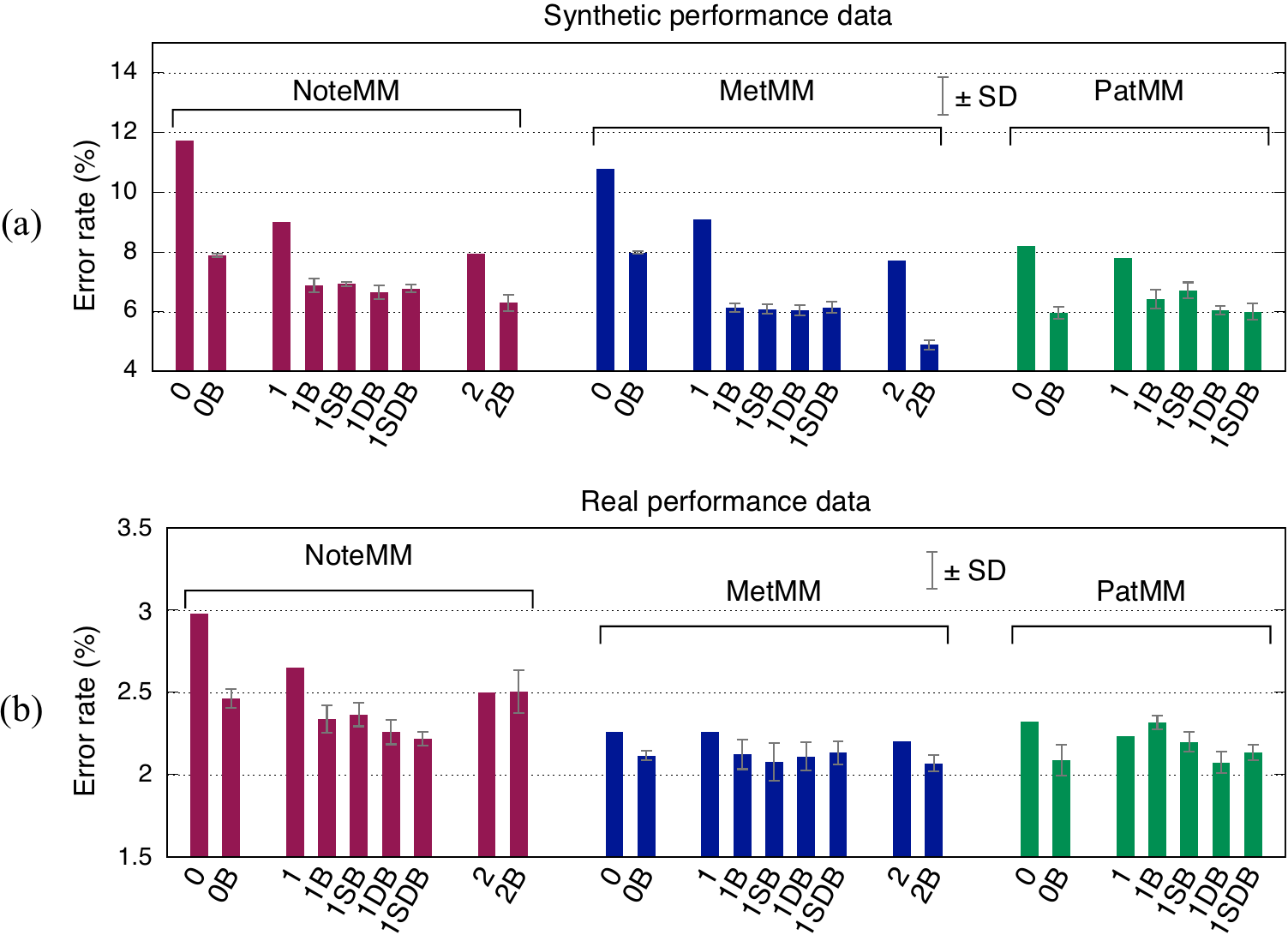}}
%\vspace{-3mm}
\caption{Transcription error rates for (a) the synthetic MIDI data and for (b) the real performed MIDI data. See Table \ref{tab:ModelList} for the model labels.}
\label{fig:ER}
\vspace{-2mm}
\end{figure}
The results are shown in Fig.~\ref{fig:ER}.
We first discuss the results for the synthetic data.
For non-Bayesian models, higher-order models outperformed the corresponding lower-order models.
The rank of the transcription accuracy approximately matches the rank of the cross entropy in Table \ref{tab:CrossEntropy}.
The Bayesian models without modification models had significantly lower error rates than the corresponding non-Bayesian models.
The effect of the Bayesian extension was often larger than that of increasing the order of the models.
For example, MetMM0B outperformed MetMM1 and MetMM1B outperformed MetMM2.
Regarding the effects of the modification models, incorporating note divisions decreased the error rate for all model types, incorporating onset shifts did so only for the metrical MM, and incorporating both operations was no more effective than incorporating only note divisions for all model types.
These results show that the incorporation of modification models can often improve the transcription results but not significantly.

Comparing the different model types, the metrical MMs consistently outperformed the corresponding note value MMs in most cases.
The accuracies of PatMM0 and PatMM1 were close to that of MetMM2, and the effect of the Bayesian extension was larger for the metrical MMs.
The best model in terms of accuracy was MetMM2B.

For the real data, most Bayesian models again significantly outperformed the corresponding non-Bayesian models.
Incorporating the modification models was effective for some cases, but there were also cases where the error rates increased.
We should notice that the error rates seem to saturate around $2.0\%$ and many models remain within the range of statistical fluctuation.
Particularly, the differences in the error rates between MetMM1 and MetMM2 and between MetMM1B and MetMM2B are much smaller than the cases with the synthetic data.
This fact also makes it difficult to clearly identify the best models.
MetMM2B again had the lowest error rate, but many other models had similar error rates.
The note value MMs were consistently worse than the corresponding metrical MMs.

Our results differ from the results of \cite{Nakamura2016} in one respect.
Paper \cite{Nakamura2016} reported a significant decrease in the error rate by incorporating the modification models into the note pattern MM.
This can be explained by the fact that the state space of the note pattern MM in \cite{Nakamura2016} did not span all possible note patterns and the search space was extended by incorporating note modifications.
This bias towards decreasing the performance of the basic note pattern MM can be clearly seen in the evaluation results (\cite{Nakamura2016}, Table 1), and it is removed in our results by constructing note pattern MMs with a search space equivalent to other types of MMs.

We manually analysed the estimation errors made by MetMM2B, which had the lowest error rates, for the real data.
There were approximately $70$ errors (corresponding to an error rate of $2.08\%$) in the transcription results and $43$ of them were clear performance errors.
Note that a performance error of one note onset usually induces two estimation errors of note values.
The remaining estimation errors were caused by less clear but significant timing deviations.
There was a performance where complex rhythms with frequent tied notes were not accurately played, and the model made $18$ estimation errors.
From these error analyses, we found that the limiting transcription accuracy of approximately $2\%$ is reasonable for this test dataset.

\begin{figure}[t]
\centering
{\includegraphics[clip,width=0.95\columnwidth]{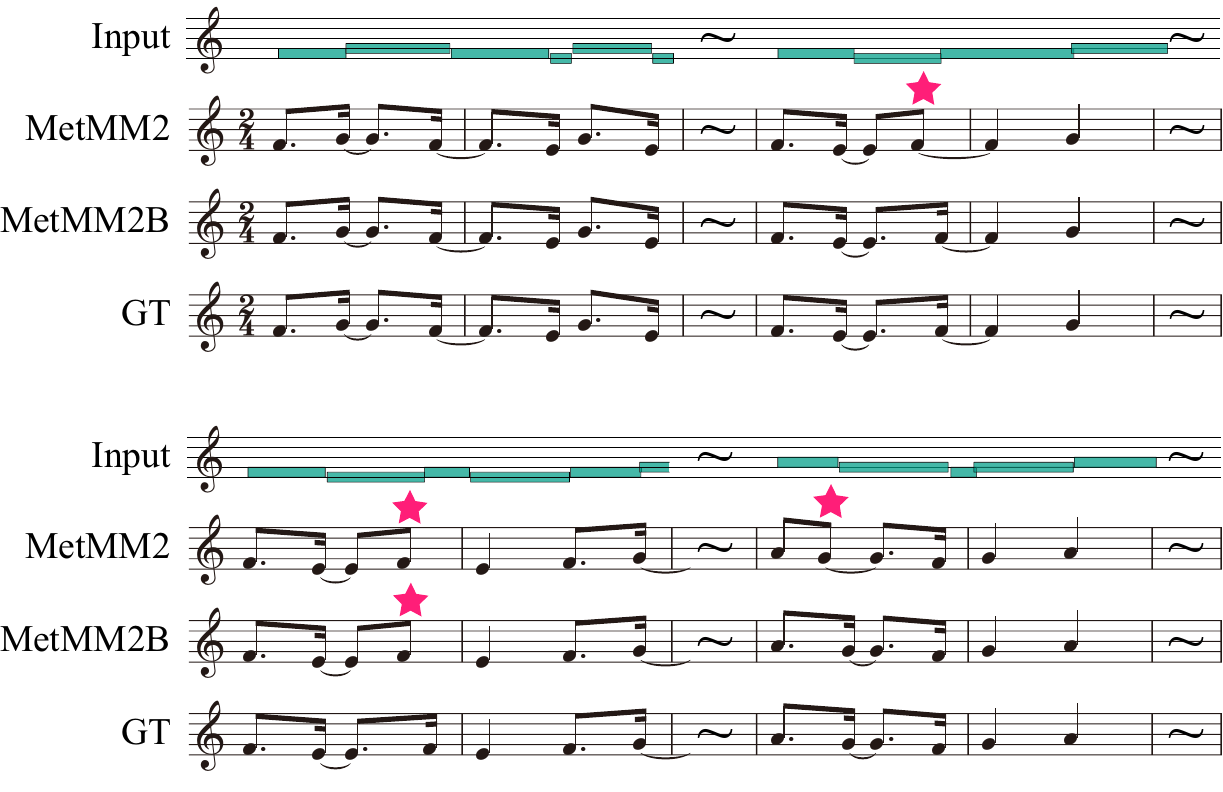}}
%\vspace{-2mm}
\caption{Examples of transcription results (RWC No.~89). Only bars with repeated rhythms, which are of our interest, are shown. GT refers to the ground truth score. Transcription errors are indicated by stars.}
\label{fig:TrEx}
%\vspace{-4mm}
\end{figure}

The transcription examples in Fig.~\ref{fig:TrEx} demonstrate the effect of the Bayesian learning of piece-specific score models.
This piece has repetitions of a tied dotted rhythm (see the first bars of the segments in Fig.~\ref{fig:TrEx}), which are played with inaccurate timings.
Since this rhythm is not common, the MetMM2 method using a generic score model made errors when the timing deviations were too large.
In contrast, the MetMM2B method was able to recognize the correct rhythm in most cases by inducing the piece-specific score model capturing these repetitions and in effect cancelling out the timing deviations.

%%%
\subsection{Influence of the Hyperparameters}
\label{sec:InfluenceHyperparameter}
%%%

%
\begin{figure}[t]
\centering
{\includegraphics[clip,width=0.9\columnwidth]{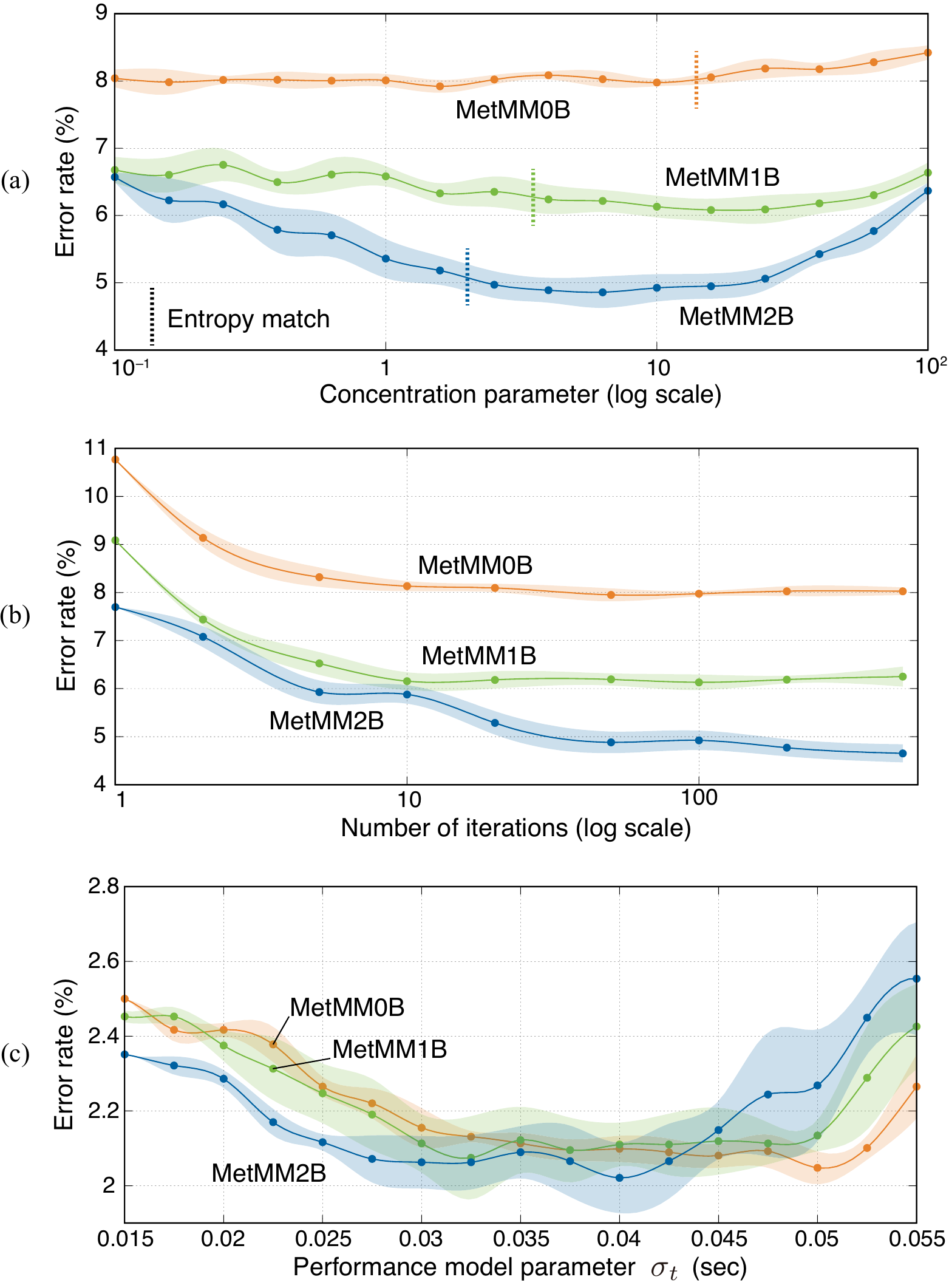}}
%\vspace{-2mm}
\caption{Transcription error rates for varying (a) the concentration parameter of the unigram/transition probabilities, (b) the number of iterations for the Gibbs sampling, and (c) the performance model parameter $\sigma_t$. The solid curves indicate the mean values and the shades indicate the ranges of $\pm1$ standard deviation.}
\label{fig:Dependence}
\vspace{-2mm}
\end{figure}
We studied the influence of the values of the hyperparameters on the transcription accuracy.
Here, we focus on the Bayesian metrical MMs that do not incorporate the modification process to investigate the generic tendencies because the non-Bayesian models have poorer performance than the Bayesian models, and the Bayesian note pattern MMs and other Bayesian models with modification models have practically prohibitive computational costs for precise measurements.

The effects of varying the values of the three most relevant parameters, including the concentration parameter for the unigram/transition probabilities, the number of iterations for Gibbs sampling, and the parameter $\sigma_t$ of the performance model, were investigated by measuring the transcription error rates as in Sec.~\ref{sec:EvaluationTranscription}.
When varying one of these parameter values, the other parameters were fixed to the values used in Sec.~\ref{sec:EvaluationTranscription}.
For the plots in Fig.~\ref{fig:Dependence}, we run each algorithm with 10 different random number seeds and obtained the average error rate.
For the concentration parameter and the number of iterations, we used the synthetic MIDI data to obtain results with small statistical fluctuations; and for performance model parameter $\sigma_t$, we used the real performed MIDI data.

Regarding the concentration parameter of the unigram/transition probabilities (Fig.~\ref{fig:Dependence}(a)), the optimal values for all the models were in the range $[1,100]$ but the values differ among the models.
The theoretically optimal values at which the expectation value of the entropies of the unigram/transition probabilities generated by the Dirichlet processes match the corresponding mean entropy of the piece-specific unigram/transition probabilities of the training data are also shown there.
We see that the theoretically optimal values do not necessarily match the empirically optimal values that maximize the transcription accuracy.
We can also confirm that the differences in the error rates for these values and the value $10$ used in Sec.~\ref{sec:EvaluationTranscription} are within the range of one standard deviation of the statistical fluctuation.

Regarding the number of iterations of the Gibbs sampling (Fig.~\ref{fig:Dependence}(b)), the error rate monotonically decreased as the number of iterations increased until a plateau was reached for each model.
For MetMM0B and MetMM1B, the plateau was already reached with approximately $10$ iterations; and for MetMM2B, the error rate seemed to continue decreasing still at the $500$\,th iteration.
Since the computation time for the parameter estimation is approximately proportional to the number of iterations, the trade off between the computation time and estimation precision is relevant, especially for higher-order models.

Regarding the performance model parameter $\sigma_t$ (Fig.~\ref{fig:Dependence}(c)), we see that different models had different optimal values in the range $[0.03,0.05]$ sec, but with a nonmonotonic behaviour and a large statistical fluctuation.
From these results, we understand that it is difficult to reliably find the precise optimal value of $\sigma_t$ for the real data and that the tentative value $0.035$ sec used in Sec.~\ref{sec:EvaluationTranscription} is at least not a bad choice.

%%%
\subsection{Discussion of the Computation Time and Extendability}
\label{sec:Discussion}
%%%

%
\begin{table*}[t]
\centering
\caption{Computation times (sec) for transcribing the real data. For the Bayesian models, the number of iterations for Gibbs sampling was $100$; and for PatMM1DB and PatMM1SDB, $W$ was $1000$. The acronyms for the models are explained in Table \ref{tab:ModelList}.}
\vspace{2mm}
{\renewcommand{\arraystretch}{1}\tabcolsep = 6pt
\begin{tabular}{lrlrlr}\toprule
Model      & Time & Model      & Time & Model      & Time \\
\midrule
NoteMM0    & $<0.1$ & NoteMM0B    & $0.13$ & NoteMM1    & $<0.1$ \\
NoteMM1B    & $0.65$ & NoteMM1SB    & $111$ & NoteMM1DB    & $23$ \\
NoteMM1SDB    & $1.4\times10^3$ & NoteMM2    & $<0.1$ & NoteMM2B    & $5.0$ \\
MetMM0     & $<0.1$ & MetMM0B     & $0.7$ & MetMM1     & $<0.1$ \\
MetMM1B     & $0.7$ & MetMM1SB     & $111$ & MetMM1DB     & $413$ \\
MetMM1SDB     & $2.0\times10^4$ & MetMM2     & $<0.1$ & MetMM2B     & $4.8$ \\
PatMM0     & $1.3$ & PatMM0B     & $558$ & PatMM1     & $1.6$ \\
PatMM1B     & $656$ & PatMM1SB     & $2.4\times10^4$ \\
PatMM1DB     & $3.3\times10^4$ & PatMM1SDB     & $9.6\times10^4$ \\
\bottomrule
\end{tabular}
}
%\vspace{2mm}
\label{tab:ComputationTime}
\vspace{-1mm}
\end{table*}
We now discuss the computation time, which is another important aspect of transcription methods.
The computation times for the studied models are listed in Table \ref{tab:ComputationTime}.
These values were measured on a computer with a 3.6 GHz Intel Core i9 CPU and 64 GB of RAM.
Roughly speaking, the computation time increases linearly as the number of notes in the input data increases and it also increases linearly as the number of iterations for Gibbs sampling for the Bayesian models increases.

From the results we see that incorporating modification models significantly increases the computation time and note pattern MMs require much more time than note value MMs and metrical MMs.
Considering both the accuracy and computation time, the first-order and second-order metrical MMs with Bayesian extensions are the most useful models in practical applications.
Incorporating modification models into these models would increase the accuracy but also significantly increase the computation time.
The order of a model, the use of the Bayesian extension and the number of iterations for Gibbs sampling, and the use of modification models should be determined according to the given resource of the computation time.

In practical applications, we need to address longer note values and finer beat resolutions than those considered in this study.
For example, if we consider the full bar length in 4/4 time and a beat resolution corresponding to one-third of a 16th note (to accommodate triplets), $N_b=48$.
The computation times required for the first-order note value MMs and metrical MMs are roughly proportional to $N_b^2$, and those for the second-order models are roughly proportional to $N_b^3$.
However, the number of possible note patterns increases as $2^{N_b}$, which makes the computation times of the note pattern MMs intractable, at least without refined searching techniques.
This means that the metrical MMs also have better extendability.

%%%%%%%%%%%%%%%%%%%%%%%%%%%%%%%%%%%%%%%%%%%%%%
\section{Conclusion}
\label{sec:Concl}
%%%%%%%%%%%%%%%%%%%%%%%%%%%%%%%%%%%%%%%%%%%%%%

We have studied the statistical description of the repetitive structure of musical notes using Bayesian score models and its application to rhythm transcription.
The main results are summarized as follows.
\begin{itemize}%\setlength{\itemindent}{-3pt}
\item The repetitive structure of musical rhythms is reflected in the sparseness of piece-specific score models, and the Dirichlet process describing the generative process of piece-specific score models from a generic score model can explain the distribution of the entropies of piece-specific models, particularly for the note value MM and metrical MM (Fig.~\ref{fig:EntropyDistribution_Other}).

\item We have confirmed that the Bayesian score models capturing repetitions were indeed effective for significantly improving the transcription accuracy.

\item Considering the transcription accuracy and the computational efficiency, the second-order Bayesian metrical MM was found to be the best among the tested models.

\item Incorporating the note modification process can help capture approximate repetitions and therefore improve the accuracy but the effect was small and the computational cost significantly increased.
\end{itemize}
In general, these results provide yet another piece of evidence for the importance of musical language modelling for music transcription and demonstrate the effectiveness of utilizing the repetitive structure.
Since repetitions are found in various musical elements including rhythms, pitch configurations, and chord progressions, the present approach can also be useful for polyphonic music transcription, chord recognition, and other automatic music recognition tasks.

As future work, it is important to extend the model and incorporate pitches and multiple voices for applications to more general types of music transcription.
The present approach of inferring piece-specific models can be even more effective in such cases that the amount of noise is larger than the case of rhythm transcription studied here.
Such a situation is common when MIDI sequences are obtained from audio signals \cite{Nakamura2018ICASSP,Nishikimi2016}.
Another interesting possibility is to combine the present Bayesian approach with deep generative models that can learn more complex structures than Markov models can learn.

%%%%%%%%%%%%%%%%%%%%%%%%%%%%%%%%%%%%%%%%%%%%%%
\section*{Supporting information}
%%%%%%%%%%%%%%%%%%%%%%%%%%%%%%%%%%%%%%%%%%%%%%

{\bf Supplemental Material.} Details of the inference algorithms for the Bayesian Markov models.

\appendix

%%%%%%%%%%%%%%%%%%%%%%%%%%%%%%%%%%%%%%%%%%%%%%
\section{Rhythm Transcription Algorithm for the First-Order Bayesian Metrical Markov Model (MetMM1B)}
\label{sec:ExplicitAlgorithm}
%%%%%%%%%%%%%%%%%%%%%%%%%%%%%%%%%%%%%%%%%%%%%%

In this appendix, an explicit algorithm for rhythm transcription based on the first-order Bayesian metrical Markov model (MetMM1B) is presented with some details of computation procedure.
We use the symbol $\mathbb{R}_+$ to represent the set of nonnegative real numbers and write $\bm a\in\mathbb{R}_+^D$ to express that $\bm a$ is a $D$-dimensional vector with nonnegative elements (probability vectors are further assumed to be normalized).

%%%
\subsection{Generative model}
%%%
The generative process of the model is summarized as follows.
The metrical Markov model (Sec.~\ref{sec:MetMM}) generates the metrical positions $b_{0:N}=(b_0,\ldots,b_N)$ as
\begin{align}
P(b_0=b)=\chi_{{\rm ini}\,b},\quad P(b_n=b\,|\,b_{n-1}=b')=\chi_{b'b}.
\end{align}
Here, each metrical position $b_n$ takes a value in $\{0,1,\ldots,N_b-1\}$ ($N_b$ is the bar length), and $\bm\chi_{\rm ini}=(\chi_{{\rm ini}\,b})_b\in\mathbb{R}_+^{N_b}$ and $\bm\chi_{b}=(\chi_{bb'})_{b'}\in\mathbb{R}_+^{N_b}$ are probability vectors.
The performance model (Sec.~\ref{sec:PerformanceModel}) generates note durations $d_{1:N}$ ($d_n>0)$ of a performance MIDI signal as
\begin{align}
P(d_n|\,b_{n-1}=b',b_n=b)=\phi_{b'bd_n}={\rm Gauss}(d_n;[b',b]v,\sigma_t^2),
\end{align}
where
\begin{align}
{\rm Gauss}(x;\mu,\sigma^2)=\frac{1}{\sqrt{2\pi\sigma^2}}\,{\rm exp}\bigg[-\frac{(x-\mu)^2}{2\sigma^2}\bigg]
\end{align}
denotes a Gaussian distribution, $v>0$ denotes the (inverse) tempo, $\sigma_t>0$ denotes the amount of onset time deviations, and $[b',b]$ represents the note value as in Eq.~(\ref{eq:BeatToScoreTime2}).
In the experimental setup of this study, parameters $v$ and $\sigma_t$ are constants specified prior to an application of the algorithm.
The complete-data probability of the whole data sequence ($d_{1:N}$ and $b_{0:N}$) is given as
\begin{align}
P(d_{1:N},b_{0:N})=\chi_{{\rm ini}\,b_0}\prod_{n=1}^N\chi_{b_{n-1}b_n}\phi_{b_{n-1}b_nd_n}.
\label{eq:CompleteDataProb}
\end{align}

The piece-specific model parameters $\bm\chi_{\rm ini}=(\chi_{{\rm ini}\,b})_b$ and $\bm\chi_{b}=(\chi_{bb'})_{b'}$ are generated from the generic model parameters $\bar{\bm\chi}_{\rm ini}\in\mathbb{R}_+^{N_b}$ and $\bar{\bm\chi}_{b}\in\mathbb{R}_+^{N_b}$ by a Dirichlet process:
\begin{align}
P(\bm\chi_{\rm ini})&={\rm Dir}(\bm\chi_{\rm ini};\alpha_{\rm ini}\bar{\bm\chi}_{\rm ini}),
\\
P(\bm\chi_{b})&={\rm Dir}(\bm\chi_{b};\alpha_{\chi}\bar{\bm\chi}_{b}),
\end{align}
where
\begin{align}
{\rm Dir}(\bm\chi;\bm\alpha)=\Gamma(\alpha_1+\cdots+\alpha_D)\prod_{i=1}^D\frac{\chi_i^{\alpha_i-1}}{\Gamma(\alpha_i)}
\end{align}
denotes a Dirichlet distribution ($\Gamma(x)$ denotes the gamma function),
% (with concentration parameter $\alpha>0$ and probability vector $\bm\chi\in\mathbb{R}_+^{D}$ and $\bm\eta\in\mathbb{R}_+^{D}$)
and $\alpha_{\rm ini}>0$ and $\alpha_{\chi}>0$ are concentration parameters.
These parameters are treated as constants to be specified prior to an application of the algorithm.

In the application of this Bayesian model for rhythm transcription, the generic model parameters $\bar{\bm\chi}_{\rm ini}$ and $\bar{\bm\chi}_{b}$ are pretrained using musical score data (Sec.~\ref{sec:ParameterLearning}).
The piece-specific model parameters $\bm\chi_{\rm ini}$ and $\bm\chi_{b}$ are treated as variables and estimated during a run of the algorithm, as described below.

%%%
\subsection{Estimation of piece-specific model parameters}
%%%

For rhythm transcription, we estimate metrical positions $b_{0:N}$ from a given performance MIDI signal with note durations $d_{1:N}$.
In the Bayesian setting, the piece-specific model parameters are estimated first, as described in this subsection, and then the metrical positions are estimated to obtain the final rhythm transcription, as described in the next subsection.
As formulated in Sec.~\ref{sec:AlgorithmForRhythmTranscription}, the former step is conducted by using the Gibbs sampling method \cite{Bishop} with the following steps of latent variable sampling and parameter sampling.

In the latent variable sampling step, model parameters $\bm\chi_{\rm ini}$ and $\bm\chi_{b}$ are fixed and $b_{0:N}$ are sampled by the forward filtering-backward sampling method.
The forward variables $F_n(b_n)=P(d_{1:n},b_n)$ are computed by the forward algorithm:
\begin{align}
F_0(b_0)&:=\chi_{{\rm ini}\,b_0}\quad(\text{initialization}),
\\
F_n(b_n)&=\sum_{b_{n-1}}\chi_{b_{n-1}b_n}\phi_{b_{n-1}b_nd_n}F_{n-1}(b_{n-1}).
\label{eq:ForwardFilteringStep}
\end{align}
The variables $b_{0:N}$ are then sampled by the backward sampling method as
\begin{align}
P(b_N|d_{1:N})&=\frac{F_N(b_N)}{\sum_bF_N(b)},
\\
P(b_{n}|\,b_{n+1:N},d_{1:N})&=\frac{\chi_{b_{n}b_{n+1}}\phi_{b_{n}b_{n+1}d_{n+1}}F_{n}(b_n)}{\sum_b\chi_{bb_{n+1}}\phi_{bb_{n+1}d_{n+1}}F_{n}(b)}.
\end{align}

In the parameter sampling step, metrical positions $b_{0:N}$ are fixed and model parameters $\bm\chi_{\rm ini}$ and $\bm\chi_{b}$ are sampled by using the probabilities
\begin{align}
P(\bm\chi_{\rm ini})&={\rm Dir}(\bm\chi_{\rm ini};\alpha_{\rm ini}\bar{\bm\chi}_{\rm ini}+\bm c_{\rm ini}),
\\
P(\bm\chi_{b'})&={\rm Dir}(\bm\chi_{b'};\alpha_\chi\bar{\bm\chi}_{b'}+\bm c^\chi_{b'}),
\end{align}
where the variables $\bm c_{\rm ini}=(c_{{\rm ini}\,b})_b$ and $\bm c^\chi_{b'}=(c^\chi_{b'b})_b$ count how many times metrical positions appear in $b_{0:N}$ and are defined as
\begin{align}
c_{{\rm ini}\,b}=\delta_{b_0b},
\quad
c^\chi_{b'b}=\sum_{n=1}^{N}\delta_{b_{n-1}b'}\delta_{b_nb}.
\end{align}
The standard method for sampling a probability vector $\bm\chi$ from a Dirichlet distribution ${\rm Dir}(\bm\chi;\bm\alpha)$ can be used: after sampling a number $y_i$ from the gamma distribution ${\rm Gam}(\alpha_i,1)$ independently for each $i$, the vector $\bm\chi$ is obtained by normalization ($\chi_i=y_i/\sum_jy_j$).

For initialization, the model parameters $\bm\chi_{\rm ini}$ and $\bm\chi_{b}$ are set to $\bar{\bm\chi}_{\rm ini}$ and $\bar{\bm\chi}_{b}$, respectively.
We then iterate the latent variable sampling step and the parameter sampling step iteratively for a predetermined number of times.
At each iteration the evidence probability $P(d_{1:N})=\sum_bF_N(b)$, calculated by Eq.~(\ref{eq:ForwardFilteringStep}) with the temporary values of $\bm\chi_{\rm ini}$ and $\bm\chi_{b}$, is memorized.
After running all iterations, the parameterization that yields the maximum evidence probability is finally chosen and used for the subsequent step for estimating metrical positions.

%%%
\subsection{Estimation of metrical positions}
%%%

After piece-specific model parameters $\bm\chi_{\rm ini}$ and $\bm\chi_{b}$ are estimated, the metrical positions $\hat{b}_{0:N}$ for the final rhythm transcription result are estimated by maximizing the probability $P(b_{0:N}|d_{1:N})$ (Sec.~\ref{sec:ParameterLearning}).
By the Bayes formula, $P(b_{0:N}|d_{1:N})\propto P(d_{1:N},b_{0:N})$ and this is equivalent to maximizing the probability in Eq.~(\ref{eq:CompleteDataProb}) w.r.t.\ $b_{0:N}$.
To solve this problem, we can apply the Viterbi algorithm \cite{Rabiner1989} as follows.
The Viterbi variables $V_n(b_n)$ and backtracking indices $U_n(b_n)$ are computed as
\begin{align}
V_0(b_0)&:=\chi_{{\rm ini}\,b_0}\quad(\text{initialization}),
\\
V_{n}(b_{n})&=\max_{b_{n-1}}\big[\chi_{b_{n-1}b_{n}}\phi_{b_{n-1}b_{n}d_{n}}V_{n-1}(b_{n-1})\big],
\\
U_{n}(b_{n})&=\argmax_{b_{n-1}}\big[\chi_{b_{n-1}b_{n}}\phi_{b_{n-1}b_{n}d_{n}}V_{n-1}(b_{n-1})\big].
\end{align}
The most probable sequence $\hat{b}_{0:N}$ is then obtained by backtracking:
\begin{align}
\hat{b}_N&=\argmax_{b_N}V_N(b_N),
\\
\hat{b}_n&=U_{n+1}(\hat{b}_{n+1}),
\end{align}
which completes our rhythm transcription algorithm.

In the actual implementation most computations involving probabilities of data sequences explained in this appendix 
are performed in the logarithmic domain to avoid overflow.
See the source code in the Supplemental Material for implementation details.
Rhythm transcription algorithms for other models studied in this paper can be derived similarly.
The details are also given in the Supplemental Material.

%\section*{References}

\end{document}

\renewcommand{\thefootnote}{\fnsymbol{footnote}}

\begin{center}
\begin{spacing}{1.5}
{\Large\bf\color{darkred}
Supplemental Material for\\``Music Transcription Based on Bayesian Piece-Specific Score Models Capturing Repetitions'' 
}
\end{spacing}

\vskip 1.2cm

Eita Nakamura$^{1,2}$\footnote[1]{Electronic address: \tt{eita.nakamura@i.kyoto-u.ac.jp}}
and Kazuyoshi Yoshii$^{1,3}$\footnote[2]{Electronic address: \tt{yoshii@kuis.kyoto-u.ac.jp}}

\vskip 0.4cm

{\it
$^1$ Graduate School of Informatics, Kyoto University, Sakyo, Kyoto 606-8501, Japan\\
$^2$ The Hakubi Center for Advanced Research, Kyoto University, Sakyo, Kyoto 606-8501, Japan\\
$^3$ AIP Center for Advanced Intelligence Project, RIKEN, Tokyo 103-0027, Japan
}

\end{center}
\vskip 1.2cm

\renewcommand{\thefootnote}{\arabic{footnote}}

\tableofcontents

%%%%%%%%%%%%%%%%%%%%%%%%%%%%%%%%%%%%%%%%%%%%%%
\section{Introduction}
\label{sec:Intro}
%%%%%%%%%%%%%%%%%%%%%%%%%%%%%%%%%%%%%%%%%%%%%%

In this Supplemental Material, we explicitly formulate three types of Markov models, their extensions incorporating note modifications, Bayesian extensions, and their integrations with the performance model.
We also explicitly describe the Gibbs sampling methods for estimating model parameters using performance data.

Based on the convention in the main text, the labels NoteMM1, NoteMM1S, NoteMM1D, and NoteMM1SD are used for the score models based on the note value Markov model (MM),  MetMM1, MetMM1S, MetMM1D, and MetMM1SD are used for the score models based on the metrical MM, and  PatMM1, PatMM1S, PatMM1D, and PatMM1SD are used for the score models based on the note pattern MM.
When these score models are integrated with the performance model, the integrated models are labelled as NoteHMM1, NoteHMM1S, etc.

%%%%%%%%%%%%%%%%%%%%%%%%%%%%%%%%%%%%%%%%%%%%%%
\section{Note Value Markov Model}
%%%%%%%%%%%%%%%%%%%%%%%%%%%%%%%%%%%%%%%%%%%%%%

%%%
\subsection{Non-Bayesian Models}
%%%

%
\subsubsection{Basic Model (NoteMM1)}

The basic note value Markov model is defined with the initial and transition probabilities:
\begin{equation}
P(r_1=r)=\pi_{{\rm ini}\,r},\quad P(r_n=r\,|\,r_{n-1}=r')=\pi_{r'r}.
\end{equation}
The complete-data probability is
\begin{align}
P(r_{1:N})=\pi_{{\rm ini}\,r_1}\prod_{n=2}^{N}\pi_{r_{n-1}r_n}.
\end{align}
\subsubsection{Model with Only Onset Shifts (NoteMM1S)}

We introduce internal variables $z_0=s_0$ and $z_n=(\tilde{r}_n,s_n)$ ($n=1,\ldots,N$), where $\tilde{r}$ indicates the note value before onset shift and $s$ indicates the amount of onset shift.
The model is defined as
\begin{align}
&P(z_0=s)=\xi_s,\quad P(z_1=(\tilde{r},s))=\pi_{{\rm ini}\,\tilde{r}}\,\xi_s,\quad
P(z_{n}=(\tilde{r},s)\,|\,z_{n-1}=(\tilde{r}',s'))=\pi_{\tilde{r}'\tilde{r}}\,\xi_s,
\\
&P(r_n|\,z_{n-1},z_n)=\mathbb{I}(r_n=\tilde{r}_n+s_n-s_{n-1}).
\end{align}
\subsubsection{Model with Only Note Divisions (NoteMM1D)}

We introduce an internal variable $z=(\tilde{r},h,g)$, where $\tilde{r}$ indicates the note value before note division and $h$ and $g$ indicate division pattern and its internal note position.
The model is defined as
\begin{align}
P(z_1=(\tilde{r},h,g))&=\pi_{{\rm ini}\,\tilde{r}}\zeta_{\tilde{r}h}\delta_{g1},
\\
P(z_{n}=(\tilde{r},h,g)\,|\,z_{n-1}=(\tilde{r}',h',g'))&=\delta_{\tilde{r}\tilde{r}'}\delta_{hh'}\delta_{g(g'+1)}+\delta_{g'G_{h'}}\pi_{\tilde{r}'\tilde{r}}\zeta_{\tilde{r}h}\delta_{g1},
\\
P(r_n|\,z_n=(\tilde{r}_n,h_n,g_n))&=\mathbb{I}(r_n=q_{h_ng_n}),
\end{align}
and the complete-data probability is
\begin{align}
&P(r_{1:N},z_{1:N})
\notag
\\
&=\pi_{{\rm ini}\,\tilde{r}_1}\zeta_{\tilde{r}_1h_1}
\delta_{g_11}\prod_{n=2}^{N}\Big(\delta_{\tilde{r}_n\tilde{r}_{n-1}}\delta_{h_nh_{n-1}}\delta_{g_n(g_{n-1}+1)}+\delta_{g_{n-1}G_{h_{n-1}}}\pi_{\tilde{r}_{n-1}\tilde{r}_n}\zeta_{\tilde{r}_nh_n}\delta_{g_n1}\Big)
\prod_{n=1}^{N}\mathbb{I}(r_n=q_{h_ng_n}).
\notag
\end{align}
\subsubsection{Model with Onset Shifts and Note Divisions (NoteMM1SD)}

We introduce internal variables $z_0=s_0$ and $z_n=(\tilde{r}_n,h_n,g_n,s_n)$ ($n=1,\ldots,N$).
The model is defined as
\begin{align}
&P(z_0=s)=\xi_s,\quad
P(z_1=(\tilde{r},h,g,s))=\pi_{{\rm ini}\,\tilde{r}}\zeta_{\tilde{r}h}\delta_{g1}\xi_s,
\\
&P(z_{n}=(\tilde{r},h,g,s)\,|\,z_{n-1}=(\tilde{r}',h',g',s'))=\big(\delta_{\tilde{r}\tilde{r}'}\delta_{hh'}\delta_{g(g'+1)}+\delta_{g'G_{h'}}\pi_{\tilde{r}'\tilde{r}}\zeta_{\tilde{r}h}\delta_{g1}\big)\xi_s,
\\
&P(r_n|\,z_{n-1},z_n)=\mathbb{I}(r_n=q_{h_ng_n}+s_n-s_{n-1}).
\end{align}
%

%%%
\subsection{Integration with the Performance Model}
%%%

%
\subsubsection{NoteHMM1}

Latent variables: $z_n=r_n$.
\begin{align}
&P(z_1=r)=\pi_{{\rm ini}\,r},\quad P(z_n=r\,|\,z_{n-1}=r')=\pi_{r'r},
\\
&P(d_n|\,z_n=r)=\phi_{rd_n}={\rm Gauss}(d_n;rv,\sigma_t^2).
\end{align}
The complete-data probability is
\begin{align}
P(d_{1:N},z_{1:N})
=\pi_{{\rm ini}\,r_1}\prod_{n=2}^{N}\pi_{r_{n-1}r_n}\prod_{n=1}^{N}\phi_{r_nd_n}.
\end{align}
\subsubsection{NoteHMM1S}

Latent variables: $z_0=s_0$, $z_n=(\tilde{r}_n,s_n)$ ($n=1,\ldots,N$).
\begin{align}
&P(z_0=s)=\xi_s,\quad
P(z_1=(\tilde{r},s))=\pi_{{\rm ini}\,\tilde{r}}\,\xi_s,\quad
P(z_{n}=(\tilde{r},s)\,|\,z_{n-1}=(\tilde{r}',s'))=\pi_{\tilde{r}'\tilde{r}}\,\xi_s,
\\
&P(d_n|z_{n-1},z_n)=\phi_{\tilde{r}_ns_{n-1}s_nd_n}={\rm Gauss}(d_n;(\tilde{r}_n+s_n-s_{n-1})v,\sigma_t^2)
\end{align}
and the complete-data probability is
\begin{align}
P(d_{1:N},z_{0:N})
=\pi_{{\rm ini}\,\tilde{r}_1}\prod_{n=2}^{N}\pi_{\tilde{r}_{n-1}\tilde{r}_n}\prod_{n=0}^{N}\xi_{s_n}\prod_{n=1}^{N}\phi_{\tilde{r}_ns_{n-1}s_nd_n}.
\end{align}
\subsubsection{NoteHMM1D}

Latent variables: $z_n=(\tilde{r}_n,h_n,g_n)$.
\begin{align}
P(z_1=(\tilde{r},h,g))&=\pi_{{\rm ini}\,\tilde{r}}\zeta_{\tilde{r}h}\delta_{g1},
\\
P(z_{n}=(\tilde{r},h,g)\,|\,z_{n-1}=(\tilde{r}',h',g'))&=\delta_{\tilde{r}\tilde{r}'}\delta_{hh'}\delta_{g(g'+1)}+\delta_{g'G_{h'}}\pi_{\tilde{r}'\tilde{r}}\zeta_{\tilde{r}h}\delta_{g1},
\\
P(d_n|\,z_n=(\tilde{r},h,g))&=\phi_{hgd_n}={\rm Gauss}(d_n;q_{hg}v,\sigma_t^2)
\end{align}
and the complete-data probability is
\begin{align}
&P(d_{1:N},z_{1:N})
\notag
\\
&=\pi_{{\rm ini}\,\tilde{r}_1}\zeta_{\tilde{r}_1h_1}
\delta_{g_11}\prod_{n=2}^{N}\Big(\delta_{\tilde{r}_n\tilde{r}_{n-1}}\delta_{h_nh_{n-1}}\delta_{g_n(g_{n-1}+1)}+\delta_{g_{n-1}G_{h_{n-1}}}\pi_{\tilde{r}_{n-1}\tilde{r}_n}\zeta_{\tilde{r}_nh_n}\delta_{g_n1}\Big)
\prod_{n=1}^{N}\phi_{h_ng_nd_n}.
\notag
\end{align}
\subsubsection{NoteHMM1DS}

Latent variables: $z_0=s_0$, $z_n=(\tilde{r}_n,h_n,g_n,s_n)$ ($n=1,\ldots,N$).
\begin{align}
&P(z_0=s)=\xi_s,\quad
P(z_1=(\tilde{r},h,g,s))=\pi_{{\rm ini}\,\tilde{r}}\zeta_{\tilde{r}h}\delta_{g1}\xi_s,
\\
&P(z_{n}=(\tilde{r},h,g,s)\,|\,z_{n-1}=(\tilde{r}',h',g',s'))=\big(\delta_{\tilde{r}\tilde{r}'}\delta_{hh'}\delta_{g(g'+1)}+\delta_{g'G_{h'}}\pi_{\tilde{r}'\tilde{r}}\zeta_{\tilde{r}h}\delta_{g1}\big)\xi_s,
\\
&P(d_n|\,z_{n-1},z_n)=\phi_{h_ng_ns_{n-1}s_nd_n}={\rm Gauss}(d_n;(q_{h_ng_n}+s_n-s_{n-1})v,\sigma_t^2),
\end{align}
and the complete-data probability is
\begin{align}
P(d_{1:N},z_{0:N})&=\pi_{{\rm ini}\,\tilde{r}_1}\zeta_{\tilde{r}_1h_1}
\delta_{g_11}\prod_{n=2}^{N}\Big(\delta_{\tilde{r}_n\tilde{r}_{n-1}}\delta_{h_nh_{n-1}}\delta_{g_n(g_{n-1}+1)}+\delta_{g_{n-1}G_{h_{n-1}}}\pi_{\tilde{r}_{n-1}\tilde{r}_n}\zeta_{\tilde{r}_nh_n}\delta_{g_n1}\Big)
\notag
\\
&\qquad\cdot\prod_{n=0}^{N}\xi_{s_n}\prod_{n=1}^{N}\phi_{h_ng_ns_{n-1}s_nd_n}.
\end{align}
%

%%%
\subsection{Bayesian Extension}
%%%

We put prior models on the parameters $\bm\pi_{\rm ini}=(\pi_{{\rm ini}\,\tilde{r}})_{\tilde{r}}$, $\bm\pi_{\tilde{r}}=(\pi_{\tilde{r}\tilde{r}'})_{\tilde{r}'}$, $\bm\zeta_{\tilde{r}}=(\zeta_{\tilde{r}h})_h$, and $\bm\xi=(\xi_s)_s$ as
\begin{align}
\bm\pi_{\rm ini}&={\rm DP}(\alpha_{\rm ini},\bar{\bm\pi}_{\rm ini}),\quad
\bm\pi_{\tilde{r}}={\rm DP}(\alpha_{\pi},\bar{\bm\pi}_{\tilde{r}}),\quad
\bm\zeta_{\tilde{r}}={\rm DP}(\alpha_{\zeta},\bar{\bm\zeta}_{\tilde{r}}),\quad
\bm\xi={\rm DP}(\alpha_{\xi},\bar{\bm\xi}).
\end{align}
%

%%%
\subsection{Gibbs Sampling for Bayesian Learning}
\label{sec:GibbsSamplingNoteMM}
%%%

Bayesian learning of the model parameters using performance data can be done with the Gibbs sampling.
Let $\theta$ be the parameters, $\bm z$ be the set of latent variables, and $\lambda$ be the collection of hyperparameters.
We sample the distribution $P(\theta,\bm z|\bm d,\lambda)$ by alternately sampling $P(\theta|\bm z,\bm d,\lambda)=P(\theta|\bm z,\lambda)$ (parameter sampling) and $P(\bm z|\theta,\bm d,\lambda)=P(\bm z|\theta,\bm d)$ (latent variables sampling).
%The computation is similar to the one in Sec.~\ref{sec:BayesianLearningForLanguageModel}.

\subsubsection{Parameter Sampling for NoteHMM1 and NoteHMM1S}

The sampling of the parameters $\theta$ from $P(\theta|\bm z,\lambda)$ can be done using the following formulas.
\begin{align}
\bm\pi_{\rm ini}&\sim{\rm Dir}(\alpha_{\rm ini}\bar{\bm\pi}_{\rm ini}+\bm c_{\rm ini}),\quad
c_{{\rm ini}\,\tilde{r}}=\delta_{\tilde{r}_1\tilde{r}},
\\
\bm\pi_{\tilde{r}'}&\sim{\rm Dir}(\alpha_\pi\bar{\bm\pi}_{\tilde{r}'}+\bm c^\pi_{\tilde{r}'}),\quad
c^\pi_{\tilde{r}'\tilde{r}}=\sum_{n=2}^{N}\delta_{\tilde{r}_{n-1}\tilde{r}'}\delta_{\tilde{r}_n\tilde{r}},
\\
\bm\xi&\sim{\rm Dir}(\alpha_\xi\bar{\bm\xi}+\bm c^\xi),\quad
c^\xi_s=\sum_{n=0}^N\delta_{s_ns}.
\end{align}
Here, $c_{{\rm ini}\,\tilde{r}}$, $c^\pi_{\tilde{r}'\tilde{r}}$, and $c^\xi_s$ count the numbers of times the relevant variables appear in the sequence $\bm z$.
Similar notation will be used in the following without mentioning.

\subsubsection{Parameter Sampling for NoteHMM1D and NoteHMM1SD}

The sampling of the parameters $\theta$ from $P(\theta|\bm z,\lambda)$ can be done using the following formulas.
\begin{align}
\bm\pi_{\rm ini}&\sim{\rm Dir}(\alpha_{\rm ini}\bar{\bm\pi}_{\rm ini}+\bm c_{\rm ini}),\quad
c_{{\rm ini}\,\tilde{r}}=\delta_{\tilde{r}_1\tilde{r}},
\\
\bm\pi_{\tilde{r}'}&\sim{\rm Dir}(\alpha_\pi\bar{\bm\pi}_{\tilde{r}'}+\bm c^\pi_{\tilde{r}'}),\quad
c^\pi_{\tilde{r}'\tilde{r}}=\sum_{n=2}^{N}\delta_{g_n1}\delta_{\tilde{r}_{n-1}\tilde{r}'}\delta_{\tilde{r}_n\tilde{r}},
\\
\bm\zeta_{\tilde{r}}&\sim{\rm Dir}(\alpha_\zeta\bar{\bm\zeta}_{\tilde{r}}+\bm c^\zeta_{\tilde{r}}),\quad
c^\zeta_{\tilde{r}h}=\sum_{n=1}^{N}\delta_{g_n1}\delta_{\tilde{r}_n\tilde{r}}\delta_{h_nh},
\\
\bm\xi&\sim{\rm Dir}(\alpha_\xi\bar{\bm\xi}+\bm c^\xi),\quad
c^\xi_s=\sum_{n=0}^N\delta_{s_ns}.
\end{align}

\subsubsection{Latent Variables Sampling}

The latent variables can be sampled using the forward filtering-backward sampling method.
The forward variable is defined as
\[
\alpha_n(z_n)=P(d_{1:n},z_n).
\]

For NoteHMM1 and NoteHMM1D, the forward variable can be computed as
\begin{align}
\alpha_1(z_1)&=P(z_1)P(d_1|z_1),
\\
\alpha_n(z_n)&=\sum_{z_{n-1}}P(d_{n}|z_n)P(z_n|z_{n-1})\alpha_{n-1}(z_{n-1}),
\\
P(\bm d=d_{1:N})&=\sum_{z_{N}}\alpha_{N}(z_{N}).
\end{align}
The backward sampling step goes as
\begin{align}
P(z_{N}|\bm d)&\propto P(\bm d,z_{N})=\alpha_{N}(z_{N}),
\\
P(z_n|\,z_{n+1:N},\bm d)&=P(z_n|\,z_{n+1},d_{1:n})
\propto P(z_n,z_{n+1},d_{1:n})=P(z_{n+1}|z_n)\alpha_n(z_n).
\end{align}

For NoteHMM1S and NoteHMM1SD, the forward variables can be computed as
\begin{align}
\alpha_0(z_0)&=P(z_0),
\\
\alpha_n(z_n)&=\sum_{z_{n-1}}P(d_{n-1}|\,z_{n-1},z_n)P(z_n|z_{n-1})\alpha_{n-1}(z_{n-1}),
\\
P(\bm d=d_{1:N})&=\sum_{z_N}\alpha_N(z_N).
\end{align}
The backward sampling step goes as
\begin{align}
P(z_N|\bm d)&\propto P(\bm d,z_N)=\alpha_N(z_N),
\\
P(z_n|\,z_{n+1:N},\bm d)&=P(z_n|\,z_{n+1},d_{1:n+1})
\propto P(z_n,z_{n+1},d_{1:n+1})=P(z_{n+1}|z_n)P(d_{n+1}|\,z_n,z_{n+1})\alpha_n(z_n).
\end{align}
%

%%%%%%%%%%%%%%%%%%%%%%%%%%%%%%%%%%%%%%%%%%%%%%
\section{Metrical Markov Model}
%%%%%%%%%%%%%%%%%%%%%%%%%%%%%%%%%%%%%%%%%%%%%%

%%%
\subsection{Non-Bayesian Models}
%%%

%
\subsubsection{Basic Model (MetMM1)}

The metrical positions $b_{0:N}$ are generated as
\[
P(b_0)=\chi_{{\rm ini}\,b_0},\quad P(b_n|b_{n-1})=\chi_{b_{n-1}b_n}.
\]
The note values and the onset score times are then given as
\[
\tau_0=b_0,\quad
r_n=\tau_{n}-\tau_{n-1}=[b_{n-1},b_n]_{N_b}=
\begin{cases}
b_{n}-b_{n-1},&b_{n-1}<b_{n};\\
b_{n}-b_{n-1}+N_b,&b_{n-1}\geq b_{n}.
\end{cases}
\]

\subsubsection{Model with Only Onset Shifts (MetMM1S)}

Introduce internal variables $z_n=(\tilde{b}_n,s_n)$ ($n=0,\ldots,N$).
\begin{align}
&P(z_0=(\tilde{b},s))=\chi_{{\rm ini}\,\tilde{b}}\xi_{s},\quad
P(z_n=(\tilde{b},s)\,|\,z_{n-1}=(\tilde{b}',s'))=\chi_{\tilde{b}'\tilde{b}}\xi_{s},
\\
&P(r_n|\,z_{n-1},z_n)=\mathbb{I}(r_n=[\tilde{b}_{n-1},\tilde{b}_n]+s_n-s_{n-1}).
\end{align}

\subsubsection{Model with Only Note Divisions (MetMM1D)}

Introduce internal variables $z_0=\tilde{b}_0$, $z_n=(\tilde{b}_n,h_n,g_n)$ ($n=1,\ldots,N$).
\begin{align}
&P(z_0=\tilde{b}_0)=\chi_{{\rm ini}\,\tilde{b}_0},\quad
P(z_1=(\tilde{b},h,g)\,|\,z_0=\tilde{b}')=\chi_{\tilde{b}'\tilde{b}}\zeta_{[\tilde{b}',\tilde{b}]h}\delta_{g1},
\\
&P(z_n=(\tilde{b},h,g)\,|\,z_{n-1}=(\tilde{b}',h',g'))=
\delta_{\tilde{b}\tilde{b}'}\delta_{hh'}\delta_{g(g'+1)}+\delta_{g'G_{h'}}\chi_{\tilde{b}'\tilde{b}}\zeta_{[\tilde{b}',\tilde{b}]h}\delta_{g1},
\\
&P(r_n|z_n)=\mathbb{I}(r_n=q_{h_ng_n}).
\end{align}

\subsubsection{Model with Onset Shifts and Note Divisions (MetMM1SD)}

Introduce internal variables $z_0=(\tilde{b}_0,s_0)$, $z_n=(\tilde{b}_n,h_n,g_n,s_n)$ ($n=1,\ldots,N$).
\begin{align}
&P(z_0=(\tilde{b},s))=\chi_{{\rm ini}\,\tilde{b}}\xi_s,\quad
P(z_1=(\tilde{b},h,g,s)\,|\,z_0=(\tilde{b}',s'))=\xi_s\chi_{\tilde{b}'\tilde{b}}\zeta_{[\tilde{b}',\tilde{b}]h}\delta_{g1},
\\
&P(z_n=(\tilde{b},h,g,s)\,|\,z_{n-1}=(\tilde{b}',h',g',s'))=
\xi_s\big(\delta_{\tilde{b}\tilde{b}'}\delta_{hh'}\delta_{g(g'+1)}+\delta_{g'G_{h'}}\chi_{\tilde{b}'\tilde{b}}\zeta_{[\tilde{b}',\tilde{b}]h}\delta_{g1}\big),
\\
&P(r_n|\,z_{n-1},z_n)=\mathbb{I}(r_n=q_{h_ng_n}+s_n-s_{n-1}).
\end{align}
%

%%%
\subsection{Integration with the Performance Model}
%%%

\subsubsection{MetHMM1}

Latent variables: $z_n=b_n$ ($n=0,\ldots,N$).
\begin{align}
&P(z_0=b)=\chi_{{\rm ini}\,b},\quad P(z_n=b\,|\,z_{n-1}=b')=\chi_{b'b},
\\
&P(d_n|\,z_{n-1}=b',z_n=b)=\phi_{b'bd_n}={\rm Gauss}(d_n;[b',b]v,\sigma_t^2),
\end{align}
and the complete-data probability is
\[
P(d_{1:N},z_{0:N})=\chi_{{\rm ini}\,b_0}\prod_{n=1}^N(\chi_{b_{n-1}b_n}\phi_{b_{n-1}b_nd_n}).
\]

\subsubsection{MetHMM1S}

Latent variables: $z_n=(\tilde{b}_n,s_n)$ ($n=0,\ldots,N$).
\begin{align}
&P(z_0=(\tilde{b},s))=\chi_{{\rm ini}\,\tilde{b}}\xi_{s},\quad
P(z_n=(\tilde{b},s)\,|\,z_{n-1}=(\tilde{b}',s'))=\chi_{\tilde{b}'\tilde{b}}\xi_{s},
\\
&P(d_n|\,z_{n-1},z_n)=\phi_{\tilde{b}_{n-1}s_{n-1}\tilde{b}_ns_nd_n}={\rm Gauss}(d_n;([\tilde{b}_{n-1},\tilde{b}_n]+s_n-s_{n-1})v,\sigma_t^2),
\end{align}
and the complete-data probability is
\begin{align}
P(d_{1:N},z_{0:N})&=\chi_{{\rm ini}\,b_0}\xi_{s_0}\prod_{n=1}^N(\chi_{\tilde{b}_{n-1}\tilde{b}_n}\xi_{s_n}\phi_{\tilde{b}_{n-1}s_{n-1}\tilde{b}_ns_nd_n}).
\end{align}

\subsubsection{MetHMM1D}

Latent variables: $z_0=\tilde{b}_0$, $z_n=(\tilde{b}_n,h_n,g_n)$ ($n=1,\ldots,N$).
\begin{align}
&P(z_0=\tilde{b}_0)=\chi_{{\rm ini}\,\tilde{b}_0},\quad
P(z_1=(\tilde{b},h,g)\,|\,z_0=\tilde{b}')=\chi_{\tilde{b}'\tilde{b}}\zeta_{[\tilde{b}',\tilde{b}]h}\delta_{g1},
\\
&P(z_n=(\tilde{b},h,g)\,|\,z_{n-1}=(\tilde{b}',h',g'))=
\delta_{\tilde{b}\tilde{b}'}\delta_{hh'}\delta_{g(g'+1)}+\delta_{g'G_{h'}}\chi_{\tilde{b}'\tilde{b}}\zeta_{[\tilde{b}',\tilde{b}]h}\delta_{g1},
\\
&P(d_n|z_n)=\phi_{h_ng_nd_n}={\rm Gauss}(d_n;q_{h_ng_n}v,\sigma_t^2),
\end{align}
and the complete-data probability is
\begin{align}
P(d_{1:N},z_{0:N})&=\chi_{{\rm ini}\,b_0}\chi_{\tilde{b}_0\tilde{b}_1}\zeta_{[\tilde{b}_0,\tilde{b}_1]h_1}\delta_{g_11}\prod_{n=1}^N\phi_{h_ng_nd_n}
\notag\\
&\quad\cdot\prod_{n=2}^N(\delta_{\tilde{b}_n\tilde{b}_{n-1}}\delta_{h_nh_{n-1}}\delta_{g_n(g_{n-1}+1)}+\delta_{g_{n-1}G_{h_{n-1}}}\chi_{\tilde{b}_{n-1}\tilde{b}_n}\zeta_{[\tilde{b}_{n-1},\tilde{b}_n]h_n}\delta_{g_n1})
.
\end{align}

\subsubsection{MetHMM1SD}

Latent variables: $z_0=(\tilde{b}_0,s_0)$, $z_n=(\tilde{b}_n,h_n,g_n,s_n)$ ($n=1,\ldots,N$).
\begin{align}
&P(z_0=(\tilde{b},s))=\chi_{{\rm ini}\,\tilde{b}}\xi_s,\quad
P(z_1=(\tilde{b},h,g,s)\,|\,z_0=(\tilde{b}',s'))=\xi_s\chi_{\tilde{b}'\tilde{b}}\zeta_{[\tilde{b}',\tilde{b}]h}\delta_{g1},
\\
&P(z_n=(\tilde{b},h,g,s)\,|\,z_{n-1}=(\tilde{b}',h',g',s'))=
\xi_s\big(\delta_{\tilde{b}\tilde{b}'}\delta_{hh'}\delta_{g(g'+1)}+\delta_{g'G_{h'}}\chi_{\tilde{b}'\tilde{b}}\zeta_{[\tilde{b}',\tilde{b}]h}\delta_{g1}\big),
\\
&P(d_n|\,z_{n-1},z_n)=\phi_{h_ng_ns_{n-1}s_nd_n}={\rm Gauss}(d_n;(q_{h_ng_n}+s_n-s_{n-1})v,\sigma_t^2),
\end{align}
and the complete-data probability is
\begin{align}
P(d_{1:N},z_{0:N})&=\chi_{{\rm ini}\,b_0}\chi_{\tilde{b}_0\tilde{b}_1}\zeta_{[\tilde{b}_0,\tilde{b}_1]h_1}\delta_{g_11}\prod_{n=0}^N\xi_{s_n}\prod_{n=1}^N\phi_{h_ng_ns_{n-1}s_nd_n}
\notag\\
&\quad\cdot\prod_{n=2}^N(\delta_{\tilde{b}_n\tilde{b}_{n-1}}\delta_{h_nh_{n-1}}\delta_{g_n(g_{n-1}+1)}+\delta_{g_{n-1}G_{h_{n-1}}}\chi_{\tilde{b}_{n-1}\tilde{b}_n}\zeta_{[\tilde{b}_{n-1},\tilde{b}_n]h_n}\delta_{g_n1}).
\end{align}
%

%%%
\subsection{Bayesian Extension}
%%%

We put prior models on the parameters as
\begin{align}
\bm\chi_{\rm ini}&={\rm DP}(\alpha_{\rm ini},\bar{\bm\chi}_{\rm ini}),\quad
\bm\chi_{\tilde{b}}={\rm DP}(\alpha_{\chi},\bar{\bm\chi}_{\tilde{b}}),\quad
\bm\zeta_{r}={\rm DP}(\alpha_{\zeta},\bar{\bm\zeta}_{r}),\quad
\bm\xi={\rm DP}(\alpha_{\xi},\bar{\bm\xi}).
\end{align}
%

%%%
\subsection{Gibbs Sampling for Bayesian Learning}
%%%

The general procedure for Bayesian learning is same as in Sec.~\ref{sec:GibbsSamplingNoteMM}.

\subsubsection{Parameter Sampling for MetHMM1 and MetHMM1S}

\begin{align}
\bm\chi_{\rm ini}&\sim{\rm Dir}(\alpha_{\rm ini}\bar{\bm\chi}_{\rm ini}+\bm c_{\rm ini}),\quad
c_{{\rm ini}\,b}=\delta_{\tilde{b}_0b},
\\
\bm\chi_{\tilde{b}'}&\sim{\rm Dir}(\alpha_\chi\bar{\bm\chi}_{\tilde{b}'}+\bm c^\chi_{\tilde{b}'}),\quad
c^\chi_{\tilde{b}'\tilde{b}}=\sum_{n=1}^{N}\delta_{\tilde{b}_{n-1}\tilde{b}'}\delta_{\tilde{b}_n\tilde{b}},
\\
\bm\xi&\sim{\rm Dir}(\alpha_\xi\bar{\bm\xi}+\bm c^\xi),\quad
c^\xi_s=\sum_{n=0}^N\delta_{s_ns}.
\end{align}

\subsubsection{Parameter Sampling for MetHMM1D and MetHMM1SD}

\begin{align}
\bm\chi_{\rm ini}&\sim{\rm Dir}(\alpha_{\rm ini}\bar{\bm\chi}_{\rm ini}+\bm c_{\rm ini}),\quad
c_{{\rm ini}\,b}=\delta_{\tilde{b}_0b},
\\
\bm\chi_{\tilde{b}'}&\sim{\rm Dir}(\alpha_\chi\bar{\bm\chi}_{\tilde{b}'}+\bm c^\chi_{\tilde{b}'}),\quad
c^\chi_{\tilde{b}'\tilde{b}}=\sum_{n=1}^{N}\delta_{g_n1}\delta_{\tilde{b}_{n-1}\tilde{b}'}\delta_{\tilde{b}_n\tilde{b}},
\\
\bm\zeta_{r}&\sim{\rm Dir}(\alpha_\zeta\bar{\bm\zeta}_{r}+\bm c^\zeta_{r}),\quad
c^\zeta_{rh}=\sum_{n=1}^{N}\delta_{g_n1}\mathbb{I}([\tilde{b}_{n-1},\tilde{b}_n]=r)\delta_{h_nh},
\\
\bm\xi&\sim{\rm Dir}(\alpha_\xi\bar{\bm\xi}+\bm c^\xi),\quad
c^\xi_s=\sum_{n=0}^N\delta_{s_ns}.
\end{align}

\subsubsection{Latent Variables Sampling}

The forward variable is defined as
\[
\alpha_n(z_n)=P(d_{1:n},z_n),
\]
which can be computed as
\begin{align}
\alpha_0(z_0)&=P(z_0),
\\
\alpha_n(z_n)&=\sum_{z_{n-1}}P(d_n|\,z_{n-1},z_n)P(z_n|z_{n-1})\alpha_{n-1}(z_{n-1}),
\\
P(\bm d=d_{1:N})&=\sum_{z_N}\alpha_N(z_N).
\end{align}

The backward sampling step goes as follows:
\begin{align}
&P(z_N|\bm d)\propto P(\bm d,z_N)=\alpha_N(z_N),
\\
&P(z_{n}|\,z_{n+1:N},\bm d)=P(z_{n}|\,z_{n+1},d_{1:n+1})
\propto P(z_{n},z_{n+1},d_{1:n+1})
\propto P(d_{n+1}|\,z_n,z_{n+1})P(z_{n+1}|z_n)\alpha_{n}(z_n).
\end{align}
%

%%%%%%%%%%%%%%%%%%%%%%%%%%%%%%%%%%%%%%%%%%%%%%
\section{Note Pattern Markov Model}
%%%%%%%%%%%%%%%%%%%%%%%%%%%%%%%%%%%%%%%%%%%%%%

%%%
\subsection{Non-Bayesian Models}
%%%

%
\subsubsection{Basic Model (PatMM1)}

The basic note pattern Markov model is described with internal variables $z_n=(k_n,i_n)$ ($n=1,\ldots,N$).
The initial and transition probabilities are given as
\begin{align}
&P(z_1=(k,i))=\psi_{{\rm ini}(ki)}=\Gamma_{{\rm ini}\,k}\gamma^{k}_{{\rm ini}\,i}=\Gamma_{{\rm ini}\,k}\delta_{i1},
\\
&P(z_n=(k,i)\,|\,z_{n-1}=(k',i'))=\psi_{(k'i')(ki)}=\delta_{kk'}\gamma^{k}_{i'i}+\gamma^{k'}_{i'\,{\rm end}}\Gamma_{k'k}\gamma^{k}_{{\rm ini}\,i}
\notag
\\
&\qquad=\delta_{kk'}\delta_{i(i'+1)}+\delta_{i'I_{k'}}\Gamma_{k'k}\delta_{i1},
\\
&P(r_n|\,z_{n-1},z_n)=\mathbb{I}(r_n=[B_{k_{n-1}i_{n-1}},B_{k_{n}i_{n}}]).
\end{align}

\subsubsection{Model with Only Onset Shifts (PatMM1S)}

Introduce internal variables $z_n=(k_n,i_n,s_n)$ ($n=0,\ldots,N$).
\begin{align}
&P(z_0=(k,i,s))=\psi_{{\rm ini}(ki)}\xi_{s},\quad
P(z_n=(k,i,s)\,|\,z_{n-1}=(k',i',s'))=\psi_{(k'i')(ki)}\xi_{s},
\\
&P(r_n|\,z_{n-1},z_n)=\mathbb{I}(r_n=[B_{k_{n-1}i_{n-1}},B_{k_{n}i_{n}}]+s_n-s_{n-1}).
\end{align}

\subsubsection{Model with Only Note Divisions (PatMM1D)}

Introduce internal variables $z_0=(k_0,i_0)$, $z_n=(k_n,i_n,h_n,g_n)$ ($n=1,\ldots,N$).
\begin{align*}
&P(z_0=(k,i))=\psi_{{\rm ini}(ki)},\quad
P(z_1=(k,i,h,g)\,|\,z_0=(k',i'))=\psi_{(k'i')(ki)}\zeta_{[B_{k'i'},B_{ki}]h}\delta_{g1},
\\
&P(z_n=(k,i,h,g)\,|\,z_{n-1}=(k',i',h',g'))=
\delta_{kk'}\delta_{ii'}\delta_{hh'}\delta_{g(g'+1)}+\delta_{g'G_{h'}}\psi_{(k'i')(ki)}\zeta_{[B_{k'i'},B_{ki}]h}\delta_{g1},
\\
&P(r_n|z_n)=\mathbb{I}(r_n=q_{h_ng_n}).
\end{align*}

\subsubsection{Model with Onset Shifts and Note Divisions (PatMM1SD)}

Introduce internal variables $z_0=(k_0,i_0,s_0)$, $z_n=(k_n,i_n,h_n,g_n,s_n)$ ($n=1,\ldots,N$).
\begin{align*}
&P(z_0=(k,i,s))=\psi_{{\rm ini}(ki)}\xi_s,\quad
P(z_1=(k,i,h,g,s)\,|\,z_0=(k',i',s'))=\psi_{(k'i')(ki)}\zeta_{[B_{k'i'},B_{ki}]h}\delta_{g1}\xi_s,
\\
&P(z_n=(k,i,h,g,s)\,|\,z_{n-1}=(k',i',h',g',s'))=
\big(\delta_{kk'}\delta_{ii'}\delta_{hh'}\delta_{g(g'+1)}+\delta_{g'G_{h'}}\psi_{(k'i')(ki)}\zeta_{[B_{k'i'},B_{ki}]h}\delta_{g1}\big)\xi_s,
\\
&P(r_n|\,z_{n-1},z_n)=\mathbb{I}(r_n=q_{h_ng_n}+s_n-s_{n-1}).
\end{align*}
%

%%%
\subsection{Integration with the Performance Model}
%%%

\subsubsection{PatHMM1}

Latent variables: $z_n=b_n$ ($n=0,\ldots,N$).
\begin{align}
&P(z_0=(k,i))=\psi_{{\rm ini}(ki)},\quad P(z_n=(k,i)\,|\,z_{n-1}=(k',i'))=\psi_{(k'i')(ki)},
\\
&P(d_n|\,z_{n-1}=(k',i'),z_n=(k,i))=\phi_{(k'i')(ki)d_n}={\rm Gauss}(d_n;[B_{k'i'},B_{ki}]v,\sigma_t^2),
\end{align}
and the complete-data probability is
\[
P(d_{1:N},z_{0:N})=\psi_{{\rm ini}(k_0i_0)}\prod_{n=1}^N(\psi_{(k_{n-1}i_{n-1})(k_ni_n)}\phi_{(k_{n-1}i_{n-1})(k_ni_n)d_n}).
\]

\subsubsection{PatHMM1S}

Latent variables: $z_n=(k_n.i_n,s_n)$ ($n=0,\ldots,N$).
\begin{align}
&P(z_0=(k,i,s))=\psi_{{\rm ini}(ki)}\xi_{s},\quad
P(z_n=(k,i,s)\,|\,z_{n-1}=(k',i',s'))=\psi_{(k'i')(ki)}\xi_{s},
\\
&P(d_n|\,z_{n-1},z_n)=\phi_{k_{n-1}i_{n-1}s_{n-1}k_ni_ns_nd_n}={\rm Gauss}(d_n;([B_{k_{n-1}i_{n-1}},B_{k_{n}i_{n}}]+s_n-s_{n-1})v,\sigma_t^2),
\end{align}
and the complete-data probability is
\begin{align}
P(d_{1:N},z_{0:N})&=\psi_{{\rm ini}(k_0i_0)}\xi_{s_0}\prod_{n=1}^N(\psi_{(k_{n-1}i_{n-1})(k_ni_n)}\xi_{s_n}\phi_{k_{n-1}i_{n-1}s_{n-1}k_ni_ns_nd_n}).
\end{align}

\subsubsection{PatHMM1D}

Latent variables: $z_0=(k_0,i_0)$, $z_n=(k_n,i_n,h_n,g_n)$ ($n=1,\ldots,N$).
\begin{align}
&P(z_0=(k_0,i_0))=\psi_{{\rm ini}(k_0i_0)},\quad
P(z_1=(k,i,h,g)\,|\,z_0=(k',i'))=\psi_{(k'i')(ki)}\zeta_{[B_{k'i'},B_{ki}]h}\delta_{g1},
\\
&P(z_n=(k,i,h,g)\,|\,z_{n-1}=(k',i',h',g'))=
\delta_{kk'}\delta_{ii'}\delta_{hh'}\delta_{g(g'+1)}+\delta_{g'G_{h'}}\psi_{(k'i')(ki)}\zeta_{[B_{k'i'},B_{ki}]h}\delta_{g1},
\\
&P(d_n|z_n)=\phi_{h_ng_nd_n}={\rm Gauss}(d_n;q_{h_ng_n}v,\sigma_t^2),
\end{align}
and the complete-data probability is
\begin{align*}
&P(d_{1:N},z_{0:N})=\psi_{{\rm ini}(k_0i_0)}\psi_{(k_0i_0)(k_1i_1)}\zeta_{[B_{k_0i_0},B_{k_1i_1}]h_1}\delta_{g_11}\prod_{n=1}^N\phi_{h_ng_nd_n}
\notag\\
&\quad\cdot\prod_{n=2}^N(\delta_{k_nk_{n-1}}\delta_{g_ng_{n-1}}\delta_{h_nh_{n-1}}\delta_{g_n(g_{n-1}+1)}+\delta_{g_{n-1}G_{h_{n-1}}}\psi_{(k_{n-1}i_{n-1})(k_ni_n)}\zeta_{[B_{k_{n-1}i_{n-1}},B_{k_ni_n}]h_n}\delta_{g_n1})
.
\end{align*}

\subsubsection{PatHMM1SD}

Latent variables: $z_0=(k_0,i_0,s_0)$, $z_n=(k_n,i_n,h_n,g_n,s_n)$ ($n=1,\ldots,N$).
\begin{align*}
&P(z_0=(k,i,s))=\psi_{{\rm ini}(ki)}\xi_s,\quad
P(z_1=(k,i,h,g,s)\,|\,z_0=(k',i',s'))=\xi_s\psi_{(k'i')(ki)}\zeta_{[B_{k'i'},B_{ki}]h}\delta_{g1},
\\
&P(z_n=(k,i,h,g,s)\,|\,z_{n-1}=(k',i',h',g',s'))=
\xi_s\big(\delta_{kk'}\delta_{ii'}\delta_{hh'}\delta_{g(g'+1)}+\delta_{g'G_{h'}}\psi_{(k'i')(ki)}\zeta_{[B_{k'i'},B_{ki}]h}\delta_{g1}\big),
\\
&P(d_n|\,z_{n-1},z_n)=\phi_{h_ng_ns_{n-1}s_nd_n}={\rm Gauss}(d_n;(q_{h_ng_n}+s_n-s_{n-1})v,\sigma_t^2),
\end{align*}
and the complete-data probability is
\begin{align*}
&P(d_{1:N},z_{0:N})=\psi_{{\rm ini}(k_0i_0)}\psi_{(k_0i_0)(k_1i_1)}\zeta_{[B_{k_0i_0},B_{k_1i_1}]h_1}\delta_{g_11}\prod_{n=0}^N\xi_{s_n}\prod_{n=1}^N\phi_{h_ng_ns_{n-1}s_nd_n}
\notag\\
&\quad\cdot\prod_{n=2}^N(\delta_{k_nk_{n-1}}\delta_{i_ni_{n-1}}\delta_{h_nh_{n-1}}\delta_{g_n(g_{n-1}+1)}+\delta_{g_{n-1}G_{h_{n-1}}}\psi_{(k_{n-1}i_{n-1})(k_ni_n)}\zeta_{[B_{k_{n-1}i_{n-1}},B_{k_ni_n}]h_n}\delta_{g_n1})
.
\end{align*}
%

%%%
\subsection{Bayesian Extension}
%%%

We put prior models on the parameters as
\begin{align}
\bm\Gamma_{\rm ini}&={\rm DP}(\alpha_{\rm ini},\bar{\bm\Gamma}_{\rm ini}),\quad
\bm\Gamma_{k}={\rm DP}(\alpha_{\Gamma},\bar{\bm\Gamma}_{k}),\quad
\bm\zeta_{r}={\rm DP}(\alpha_{\zeta},\bar{\bm\zeta}_{r}),\quad
\bm\xi={\rm DP}(\alpha_{\xi},\bar{\bm\xi}).
\end{align}
%

%%%
\subsection{Gibbs Sampling for Bayesian Learning}
%%%

The general procedure for Bayesian learning is same as in Sec.~\ref{sec:GibbsSamplingNoteMM}.

\subsubsection{Parameter Sampling for PatHMM1 and PatHMM1S}

\begin{align}
\bm\Gamma_{\rm ini}&\sim{\rm Dir}(\alpha_{\rm ini}\bar{\bm\chi}_{\rm ini}+\bm c_{\rm ini}),\quad
c_{{\rm ini}\,k}=\delta_{k_0k},
\\
\bm\Gamma_{k'}&\sim{\rm Dir}(\alpha_\Gamma\bar{\bm\Gamma}_{k'}+\bm c^\Gamma_{k'}),\quad
c^\Gamma_{k'k}=\sum_{n=1}^{N}\delta_{i_n1}\delta_{k_{n-1}k'}\delta_{k_nk},
\\
\bm\xi&\sim{\rm Dir}(\alpha_\xi\bar{\bm\xi}+\bm c^\xi),\quad
c^\xi_s=\sum_{n=0}^N\delta_{s_ns}.
\end{align}

\subsubsection{Parameter Sampling for PatHMM1D and PatHMM1SD}

\begin{align}
\bm\Gamma_{\rm ini}&\sim{\rm Dir}(\alpha_{\rm ini}\bar{\bm\chi}_{\rm ini}+\bm c_{\rm ini}),\quad
c_{{\rm ini}\,k}=\delta_{k_0k},
\\
\bm\Gamma_{k'}&\sim{\rm Dir}(\alpha_\Gamma\bar{\bm\Gamma}_{k'}+\bm c^\Gamma_{k'}),\quad
c^\Gamma_{k'k}=\sum_{n=1}^{N}\delta_{g_n1}\delta_{i_n1}\delta_{k_{n-1}k'}\delta_{k_nk},
\\
\bm\zeta_{r}&\sim{\rm Dir}(\alpha_\zeta\bar{\bm\zeta}_{r}+\bm c^\zeta_{r}),\quad
c^\zeta_{rh}=\sum_{n=1}^{N}\delta_{g_n1}\mathbb{I}([B_{k_{n-1}i_{n-1}},B_{k_ni_n}]=r)\delta_{h_nh},
\\
\bm\xi&\sim{\rm Dir}(\alpha_\xi\bar{\bm\xi}+\bm c^\xi),\quad
c^\xi_s=\sum_{n=0}^N\delta_{s_ns}.
\end{align}

\subsubsection{Latent Variables Sampling}

The forward variable is defined as
\[
\alpha_n(z_n)=P(d_{1:n},z_n),
\]
which can be computed as
\begin{align}
\alpha_0(z_0)&=P(z_0),
\\
\alpha_n(z_n)&=\sum_{z_{n-1}}P(d_n|\,z_{n-1},z_n)P(z_n|z_{n-1})\alpha_{n-1}(z_{n-1}),
\\
P(\bm d=d_{1:N})&=\sum_{z_N}\alpha_N(z_N).
\end{align}

The backward sampling step goes as follows:
\begin{align*}
&P(z_N|\bm d)\propto P(\bm d,z_N)=\alpha_N(z_N),
\\
&P(z_{n}|\,z_{n+1:N},\bm d)=P(z_{n}|\,z_{n+1},d_{1:n+1})
\propto P(z_{n},z_{n+1},d_{1:n+1})
\propto P(d_{n+1}|\,z_n,z_{n+1})P(z_{n+1}|z_n)\alpha_{n}(z_n).
\end{align*}